\documentclass[prb,aps,twocolumn,groupaddress,nofootinbib,floatfix]{revtex4-1}
\usepackage{latexsym}

\usepackage{graphicx}
\usepackage{verbatim}
\usepackage{multirow}
\usepackage{amsmath}
\usepackage{mathrsfs}
\usepackage{float}
\usepackage[usenames, dvipsnames]{color}
\usepackage{mathtools}
\usepackage{pgfplots}
\pgfplotsset{compat=1.4}
\usepgfplotslibrary{groupplots}

\usepackage{slashed}
\usepackage{physics}	
\usepackage{graphicx}   
\usepackage{epstopdf}
\usepackage{bbold}
\usepackage{wasysym}

\usepackage[ampersand]{easylist}
\let\eval\undefined
\newcommand{\beq}{\begin{eqnarray}}
\newcommand{\eeq}{\end{eqnarray}}
\newcommand \eval[1] {\Bigr|_{#1}}
\newcommand \adj     {^\dagger}
\newcommand \inv     {^{-1}}

\renewcommand \bar   [1] {\overline{#1}}
\newcommand \half   {\frac{1}{2}}

\newcommand \NN     {\nonumber}
\renewcommand \P     [1]{\left(#1\right)}
\newcommand   \B     [1]{\left[#1\right]}

\newcommand   \ave   [1]{\langle{#1}\rangle}

\newcommand   \<       {\langle}
\renewcommand \>       {\rangle}
\newcommand   \proj[1] {\lvert #1 \>\< #1 \rvert}

\newcommand   \qave[3] {\< #1 | #2 | #3 \>}
\newcommand \cH {\mathcal{H}}

\newcommand   \ox  {\otimes}
\newcommand   \oo  {\infty}

\renewcommand   \tr      {\operatorname*{tr}}

\newcommand \GS {{\mathrm{GS}}}
\newcommand \HK {{\mathrm{HK}}}
\newcommand \Id {\mathbb{1}}
\newcommand \up {\uparrow}
\newcommand \dn {\downarrow}

\usepackage[final,unicode,colorlinks=true,citecolor=purple,linkcolor=blue,linktocpage]{hyperref}
\usepackage[obeyFinal,textsize=tiny,backgroundcolor=white,linecolor=magenta]{todonotes}

\begin{document}

\title{Exact Superconducting Instability in a Doped Mott Insulator}

\author{Philip W. Phillips}
\author{Luke Yeo}
\author{Edwin W. Huang}
\affiliation{Department of Physics and Institute of Condensed Matter Theory, University of Illinois at Urbana-Champaign, Urbana, 61801 Illinois, USA}

\begin{abstract}
Because the cuprate superconductors are doped Mott insulators, it would be advantageous to solve even a toy model that exhibits both Mottness and superconductivity. We consider the Hatsugai-Kohmoto model~\cite{hk1992,baskaran1991exactly}, an exactly solvable system that is a prototypical Mott insulator above a critical interaction strength at half filling. Upon doping or reducing the interaction strength, our exact calculations show that the system becomes a non-Fermi liquid metal with a superconducting instability. In the presence of a weak pairing interaction, the instability produces a thermal transition to a superconducting phase, which is distinct from the BCS state, as evidenced by a gap-to-transition temperature ratio exceeding the universal BCS limit. The elementary excitations of this superconductor are not Bogoliubov quasiparticles but rather superpositions of doublons and holons, composite excitations signaling that the superconducting ground state of the doped Mott insulator inherits the non-Fermi liquid character of the normal state. An unexpected feature of this model is that it exhibits a superconductivity-induced transfer of spectral weight from high to low energies as seen in the cuprates\cite{marel1} as well as a suppression of the superfluid density relative to that in BCS theory.
\end{abstract}

\maketitle

Cooper's~\cite{cooper} demonstration that the normal state of a metal is unstable to a pairing interaction between two electrons above the Fermi surface paved the way to the eventual solution to the problem of superconductivity in elemental metals such as mercury.  In modern language~\cite{shankar,polchinski,benfatto}, the Cooper instability is understood as the only relevant perturbation along a Fermi surface given that all renormalizations due to short-ranged repulsive interactions are benign.  The problem of high-temperature superconductivity in the copper-oxide ceramics persists because the normal state is a doped Mott insulator in which no organizing principle such as quasiparticles on a Fermi surface can be invoked.  The question arises: Is there an analogue of Cooper's argument for a doped Mott insulator?  Such a demonstration would be non-trivial as the simplest model relevant to the cuprates, namely the Hubbard model, is intractable in $d>1$.  Given this intractability, we seek a simplification.  Namely, is there a simplified model which captures key features of Mottness but still permits a definitive answer to the Cooper problem?

We demonstrate such an instability for the  Hatsugai-Kohmoto (HK)~\cite{hk1992,baskaran1991exactly} model of a doped Mott insulator.   The minimum feature of Mottness~\cite{RMP,dzy}, thereby setting it apart from a Fermi liquid, is a bifurcation of the spectral weight per momentum state into low and high-energy components.  Such a bifurcation creates a surface of zeros of the single-particle Green function, connoted a Luttinger surface~\cite{dzy}, and is known to be essential to describing high-low energy mixing in doped Mott systems~\cite{marel1,dzy,sawatzky,eskes,chen,uchida,RMP,bontemps}.  As the HK model is the simplest example which captures how quasiparticles (poles of the single-particle Green function) on a Fermi surface are converted to zeros, any superconducting instability found in such a setup could ultimately illuminate the solution to the full problem.  Recent progress, in fact part of the motivation for this paper, has been made along these lines  in the zero chemical potential limit\cite{chandan} in a phenomenological model for the Luttinger surface. 

A key experimental signature of superconductivity in the cuprates that any model of pairing in a Mott insulator should inform is the in-plane transfer of spectral weight from high to low energies, dubbed the color change\cite{marel1}.  Specifically the integrated weight of the optical conductivity over the lower Hubbard band scales ($<1eV$) increases below the superconducting temperature, whereas the high-energy component ($[1eV,2eV]$) decreases.  Since the integrated weight determines the number of charge carriers, the color change indicates that high-energy scales contribute to the superfluid density in contrast to the standard BCS picture.  In addition, the superfluid density is suppressed relative to its value in BCS theory\cite{bozovic}.  We show that the HK model exhibits both of these features as a direct consequence of Mottness.   The latter arises from just the splitting of the spectrum into lower and upper bands, while the first  is a consequence of dynamical mixing between the upper and lower Hubbard bands induced by the pairing interaction.  As dynamical (hopping-driven) mixing between between the upper and lower Hubbard bands is present in the Hubbard model as a result of the non-commutativity ~\cite{sawatzky,RMP,chen,eskes} of the kinetic and potential energy terms, our conclusion that dynamical spectral weight transfer (DSWT) is the mechanism for the color change transcends the HK model and is a general consequence of Mottness.

Because the HK model is a simplification of the Hubbard model, a loose analogy (based on the all-to-all interactions) with the Sachdev-Ye-Kitaev (SYK)~\cite{sy1993,k2015} model  applied to the strange metal phase of the cuprates is appropriate  as while they do not mirror the physics accurately, they do offer controlled analytics on non-Fermi liquid states.
The HK model is probably more powerful in this regard as it actually models a Mott insulator with a Luttinger surface that gives rise to a non-Fermi liquid upon doping. What the HK model lays plain is that DSWT can be separated from the presence of the Mott gap. With this in mind, we perform calculations in the non-Fermi liquid state and study the superconducting instability through an exact calculation of the pair-field susceptibility.  We then include a weak pairing interaction and explore the nature of the superconducting ground state and its elementary excitations, finding fundamental differences with the BCS ground state that ultimately arise from the non-Fermi liquid nature of the doped Mott insulator.

As in the SYK model~\cite{sy1993}, a key ingredient that makes the HK and Baskaran\cite{baskaran1991exactly} models tractable is the presence of all-to-all interactions.  In the HK model and the Baskaran model\cite{baskaran1991exactly} as well,
\begin{eqnarray}
H_{\rm HK} &=& -t \sum_{\langle j,l\rangle,\sigma} \pqty{ c^\dagger_{j\sigma} c^{}_{l\sigma} + h.c. }  - \mu \sum_{j\sigma} c^\dagger_{j\sigma} c^{}_{j\sigma}\\
&& + \frac{U}{L^d} \sum_{j_1..j_4}\delta_{j_1+ j_3, j_2+ j_4} c^\dagger_{j_1\uparrow} c^{}_{j_2\uparrow} c^\dagger_{j_3\downarrow} c^{}_{j_4\downarrow},
\end{eqnarray}
the interaction term is not random but a constant, $U$, and unlike SYK, a hopping term ($t$) is present between nearest neighbors $\langle j,l\rangle$ that gives the model dimensionality.
An additional feature is the presence of a constraint $j_1 + j_3 = j_2 + j_4$ that the electrons must satisfy for the interaction term $U$ to be felt.
Here $\mu$ is the chemical potential and $L^d$ is the number of lattice sites.
While the SYK model is tractable only in the limit of a large number of flavors, the HK model is exactly solvable as can be seen from Fourier transforming to momentum space
\begin{equation}
	H_{\rm HK} = \sum_{k}H_{k} = \sum_{k} \left (\xi_k(n_{k\uparrow} + n_{k\downarrow} ) + U  n_{k\uparrow} \, n_{k\downarrow}\right).
\label{eq:kSpaceHK}
\end{equation}
Here $n_{k\sigma} = c_{k\sigma}\adj c_{k\sigma}$ is the fermion number operator for the mode with momentum $k$ and spin $\sigma=\up,\dn$. It is clear that the kinetic and potential energy terms commute.  Consequently, momentum is a good quantum number, unlike the Hubbard model, and all eigenstates have a fixed unfluctuating occupancy in $k$-space. Here, the momenta are summed over a square Brillouin zone $[-\pi, \pi)^d$, within which the quasiparticle spectrum $\xi_k = \epsilon_k - \mu$ is set by the dispersion $\epsilon_k = -(W/2d) \sum_{\mu=1}^{d} \cos k^\mu$ with non-interacting bandwidth $W = 4 d t$ and offset by a chemical potential $\mu$.

\begin{figure}
\centering
\includegraphics{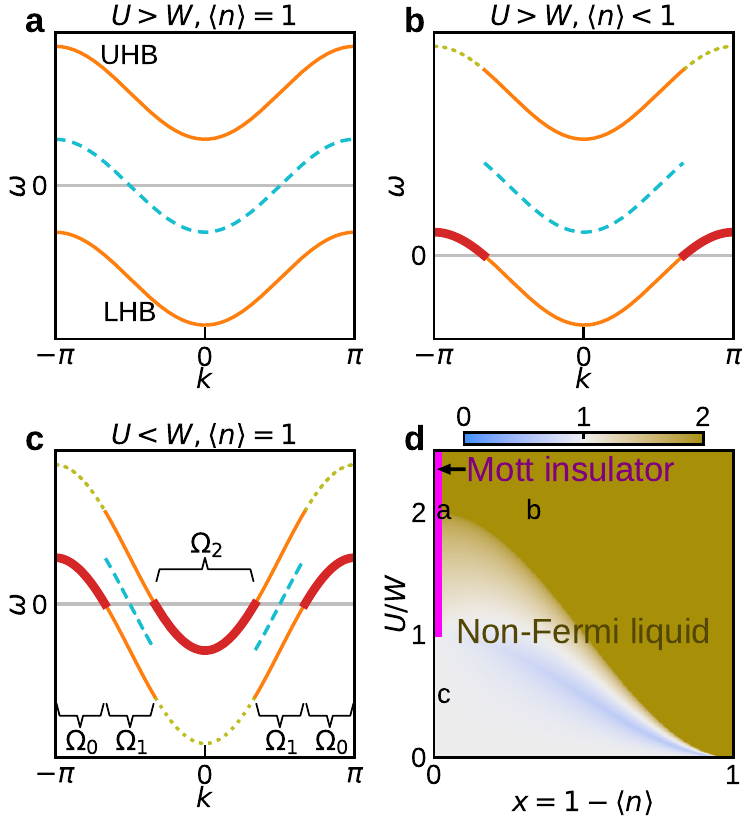}
\caption{{\bf Single-particle Green functions and phase diagram of the HK model.} {\bf a} - {\bf c}, Poles and zeros of the single-particle Green function (Eq. \ref{eq:propagator}). A 1d tight-binding dispersion is used for simplicity. Zeros are indicated by dashed blue lines. Poles with weight $0$, $0.5$, and $1$ are indicated by dotted olive, thin orange lines, and bold red lines respectively. Upper Hubbard band (UHB) and lower Hubbard band (LHB) are labelled in {\bf a}. Regions of occupancy $\ev{n_k} = i$ are labelled as $\Omega_i$ for $i=0,1,2$ in {\bf c}. {\bf d}, Ground state phase diagram of the HK model. The non-Fermi liquid covers the entire diagram except for the half-filled Mott insulator for $U>W$. Color represents the ratio of Luttinger count to filling; deviation from $1$ (white) indicates violation of Luttinger's theorem.}
\label{fig:phase-diagram}
\end{figure}

What is surprising about the HK model is that although the potential and kinetic energy terms commute, a correlated metal-insulator transition still exists~\cite{hk1992}.
To shed new light on HK physics and the excitation spectrum, we focus on the structure of the single-particle Green function
\begin{align}\label{eq:propagator}
    G_{k\sigma}(i\omega_n)
    &\equiv -\int_{0}^{\beta} d\tau\; \ave{c_{k\sigma}(\tau) c_{k\sigma}\adj(0)} e^{i\omega_n\tau} \\
    G_{k\sigma}(i\omega_n \to z) &= \frac{1-\ave{n_{k\bar{\sigma}}}}{z - \xi_k} + \frac{\ave{n_{k\bar{\sigma}}}}{z - (\xi_k + U)}, 
\end{align}
which is plotted in Fig.~\ref{fig:phase-diagram}a-c. The two-pole structure is reminiscent of the atomic limit of the Hubbard model except now $i\omega_n$ is replaced with $i\omega_n-\xi_k$.  The corresponding density of states shares a mutual energy region only for $U<W$.  Consequently, a gap ($\Delta E=U-W$) appears in the single-particle spectrum for $U>W$ resulting in a Mott insulating state at half-filling (Fig.~\ref{fig:phase-diagram}a), with $\langle n_{k\bar\sigma} \rangle=1/2$ for all $k$. Doping away from half-filling or reducing interaction strength $U<W$, the gap shifts away from the chemical potential or vanishes entirely, leading to a compressible metallic state (Fig.~\ref{fig:phase-diagram}b and c). In this metallic state, the momentum occupancy changes discontinuously from $\ev{n_k} = \ev{n_{k\uparrow} + n_{k\downarrow}}=1$ for singly occupied momenta in $\Omega_1$ to $2$ for the doubly occupied part ($\Omega_2$) and $0$ for the empty region ($\Omega_0$).

Spin-rotation invariance of the HK model dictates that although singly-occupied momenta exist in the metal, they cannot appear as pure states.  The metal is the mixed state consisting of a uniform ensemble over all spin states of the form
\begin{eqnarray}
|\Psi_G;\{\sigma_k\}\rangle=\prod_{k \in \Omega_1}c^\dagger_{k\sigma_k} \prod_{k \in \Omega_2} c_{k\uparrow}^\dagger c_{k\downarrow}^\dagger|0\rangle, \label{eq:HKGS}
\end{eqnarray}
which results in a large ground-state degeneracy. Consequently, the excitations in the metallic state have no well-defined spin.  In fact, they do not even have well defined charge. Because of the lack of mixing between the two Hubbard bands, excitations in the lower band  are created by $\zeta_{k\sigma}=c^\dagger_{k\sigma}(1-n_{k\bar\sigma})$ at energy $\xi_k$ while those in the upper by $\eta_{k\sigma}=c_{k\sigma}^\dagger n_{k\bar\sigma}$ at energy $\xi_k + U$.  As a result, the  excitations in the metal are more akin to doublon and holon composite excitations and hence the metal of the HK model lacks any interpretation in terms of Fermi-liquid quasiparticles.  This can be seen directly from the retarded Green function
\beq
    G^R_\sigma(k,\omega)
    = \frac{1}{\displaystyle \omega + i0^+ - (\xi_k+U/2) - \frac{(U/2)^2}{\omega + i0^+ - (\xi_k+U/2)}}
\eeq
when the occupancies are equal.  Both real and imaginary parts of the self-energy diverge at $\omega=0$ along the surface defined by $\xi_k=-U/2$.  Such a divergence indicates that the Green function cannot be derived perturbatively from the non-interacting limit; there is no Fermi liquid.

The two-pole structure determines the sign-changes of the Green function and is crucial in calculating the Luttinger count $2 \sum_k \theta(\Re G(k, 0))$. Except for fine-tuned cases, such as at $\mu=U/2$ where there is exact particle-hole symmetry, or when $U=0$ where the model is noninteracting, Luttinger's theorem is violated in the HK model and Hubbard models in similar limits\cite{dave2013,rosch}. The Luttinger count is decoupled from the true occupancy and can exceed it by up to a factor of $2$ (Fig.~\ref{fig:phase-diagram}d). This is evident deep in the doped Mott insulating regime (Fig.~\ref{fig:phase-diagram}b), where $\Omega_1$ is singly occupied but contributes fully to the Luttinger count when the zeros of the Green function are above the chemical potential. The violation of Luttinger's theorem throughout the phase diagram (Fig.~\ref{fig:phase-diagram}d) indicates that the metallic state of the HK model is incompatible with Fermi-liquid theory. 

Having established that we have a completely controlled non-Fermi liquid metallic state, we can address the question:  is such a state unstable to pairing?   To this end, we append the HK Hamiltonian with an attractive ($g>0$) pairing interaction,
\begin{eqnarray}
H = H_\HK - g H_p,
\qquad H_p = \frac{1}{L^d} \Delta\adj \Delta
\label{Htot}
\end{eqnarray}
where $\Delta = \sum_k b_k= \sum_k c_{-k\downarrow} c_{k\uparrow}$ is the $s$-wave pair creation operator at zero total momentum.
Seeking an analogue of Cooper's argument, we first focus on the pair amplitude
\begin{eqnarray}
i\hbar{\partial\over \partial t}\alpha_{k}
(t=0)&=&i\hbar{\partial\over \partial t}\langle\GS(t)\left|b_{k}\right|\psi(t)\rangle\nonumber\\ &=& \langle
\GS(t)\left|\left[b_{k},H\right]\right|\psi(t)\rangle,
\end{eqnarray}
where $|\rm GS\rangle$ is a metallic state in the zero-temperature ensemble described by Eq.~(\ref{eq:HKGS}), and $|\psi\rangle$ is the state with a single pair given by
\begin{eqnarray}
|\psi\rangle=\sum_{k\in\Omega_0}\alpha_k b^\dagger_k|\rm GS\rangle.
\end{eqnarray}
For clarity, we take a maximally polarised state for $\ket{\GS} = \P{ \prod_{k \in \Omega_2} c_{k\up}\adj c_{k\dn}\adj } \P{ \prod_{k \in \Omega_1} c_{k\up}\adj } \ket{0}$.
(In the supplementary text~\cite{supp} we show that the zero temperature Gibbs state recovers the same result.)
From
\begin{align}
    [b_k,H]
    &= \P{ 2\xi_k + U(n_{k\downarrow}+n_{-k\uparrow}) } b_k \NN\\
    &\quad- \frac{g}{L^d}(1-n_{k\uparrow}-n_{-k\downarrow}) \sum_{k'} b_{k'},
\end{align}
the equations of motion take on the form
\begin{align}
    &\quad (i\hbar\partial_t - 2\xi_k - U \ave{n_{k\dn} + n_{-k\up}}) \alpha_k \NN\\
    &= -\frac{g}{L^d} \ave{1 - n_{k\up} - n_{-k\dn}} \sum_{k'}\alpha_{k'}
\end{align}
where $\langle\cdots\rangle$ denotes an expectation value in the state $|\rm GS\rangle$. A similar equation can be derived for the Baskaran\cite{bask} model as well.

\begin{figure}
    \centering
    \includegraphics{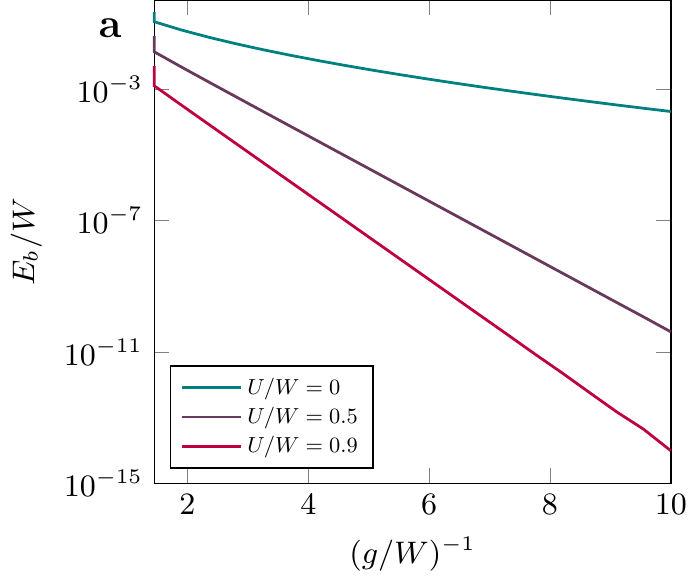}
    \\\vspace{1em}
    \includegraphics{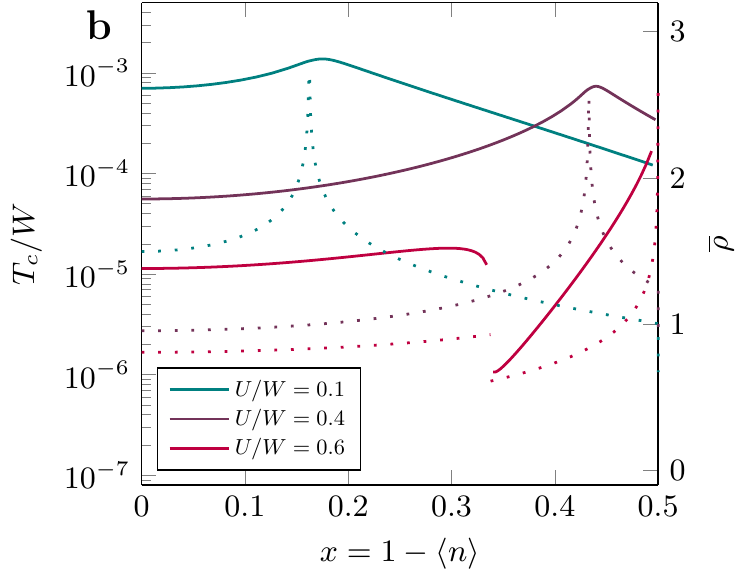}
    \caption{{\bf Superconducting energy scales in two dimensions.}
        {\bf a}, Cooper pair binding energy $E_b$ at half-filling ($x=0$).
        The linear regime $\ln (E_b/W) \sim -(g/W)\inv$ is achieved at large values of the inverse pair coupling $(g/W)\inv$.
        {\bf b}, Superconducting temperature $T_c$ (solid) at pair coupling $g/W = 0.1$, over a range of hole dopings $x = 1 - \sum_{k,\sigma} \ave{n_{k\sigma}}/L^d$ away from half-filling.
        Its qualitative behavior under doping is given by the mean density of states $\bar{\rho} = \half(\rho(\mu) + \rho(\mu-U))$ (dotted) at the same pair coupling.}
    \label{fig:EbTc}
\end{figure}

We solve this equation in the standard way by letting $\alpha_k(t)=e^{-iEt/\hbar}\alpha_k(t=0)$.
Dividing  by the coefficient on the left-hand side and performing the sum over momentum, we obtain
\begin{eqnarray}
    1=-\frac{g}{L^d}\sum_{k\in\Omega_0}\frac{\ave{1-n_{k\up}+n_{-k\dn}}}{E-2\xi_k-U\ave{n_{k\dn} + n_{-k\up}}}
\end{eqnarray}
as the familiar criterion for a superconducting instability. In the range of integration, $\ave{n_{k\sigma}}=0$, resulting in the simplified expression,
\begin{eqnarray}
1=-g\int_{\mu}^{W/2} d\epsilon \frac{\rho(\epsilon)}{E-2\epsilon+2\mu},
\end{eqnarray}
having converted the sum to an integral weighted by the density of states $\rho(\epsilon)$ for the band $\epsilon_k$.
This is, up to the limits of integration and density of states, exactly of the BCS form.
It therefore results in a bound-state energy $E<0$ for any $g>0$, which we plot in Fig.~\ref{fig:EbTc} for $d=2$.
In the case of half-filled metal ($\mu=U/2$) in one dimension, for example, the binding energy
\begin{equation}\label{eq:binding}
    E_b = -E \sim W (1-(U/W)^2) e^{-\pi W \sqrt{1-(U/W)^2}/g}
\end{equation}
is exponentially small in $1/g$.
The full $d$-dimensional dependence is shown in the supplementary materials~\cite{supp}.  Hence, the HK model exhibits an analogue of the instability Cooper\cite{cooper} found for a Fermi liquid and answers the question affirmatively asked in Ref.\cite{mross}.

The advantage of the HK model is that we need not settle on the pair-binding calculation to determine whether a superconducting instability exists.  We can compute the pair susceptibility
\begin{equation}
    \chi(i\nu_n) \equiv \frac{1}{L^d} \int_{0}^{\beta} d\tau\; e^{i\nu_n \tau} \ave{T \Delta(\tau) \Delta\adj}_g
\end{equation}
exactly at all temperatures, though below we will emphasize the low temperature regime $T \ll U,W$.
As we show in the supplementary text~\cite{supp}, $\chi(i\nu_n)$ is related to the `bare' susceptibility, $\chi_0(i\nu_n)$ at $g=0$, through the Dyson equation
\begin{equation}
    \chi = \chi_0 + g \chi_0 \chi ,
\end{equation}
where $\ave{\cdots}_g$ (resp.\ $\ave{\cdots}_0$) is an expectation value in the Gibbs state $e^{-\beta H}/Z$ (resp.\ $e^{-\beta H_\HK}/Z_0$), and $\nu_n = 2\pi n/\beta$ is a bosonic Matsubara frequency.
Then the fluctuation propagator $L \equiv -g \chi/\chi_0$ satisfies the usual equation~\cite{varlamov-galda-glatz}
\begin{equation}
L = -g + g \chi_0 L = \frac{1}{\chi_0 - 1/g}
\end{equation}
such that $L(\omega=0)$ diverges when $\chi_0(0) \eval{T=T_c} = 1/g$, thereby fixing the critical superconducting temperature $T_c$.
In order to compute $\chi_0$, we first simplify
\begin{align}
    \ave{T \Delta(\tau) \Delta\adj}_0
    &= \sum_{k,p} \ave{T c_{-k\dn}(\tau) c_{k\up}(\tau) c_{p\up}\adj c_{-p\dn}\adj}_0 \\
    &= \sum_{k} \ave{T c_{-k\dn}(\tau) c_{k\up}(\tau) c_{k\up}\adj c_{-k\dn}\adj}_0 \\
    &\sim \sum_{k} G_{-k\dn}(\tau) G_{k\up}(\tau), \label{eq:bare-factorization}
\end{align}
up to an unimportant subextensive contribution from coincident terms with $k=-k$, having used that the ensemble consists only of Fock states (guaranteed by the preservation of $n_{k\sigma}$ as a good quantum number in the HK model) in the second line, and in the third line that the Gibbs state $e^{-\beta H_\HK}/Z_0$ factorizes in $k$-space.  We note that despite the appearance of Eq.~(\ref{eq:bare-factorization}), we have not utilized Wick's theorem, which does not apply in general to the HK model.  
Writing out the single-particle Green function,
\begin{equation}
    -G_{k\sigma}(\tau) = \ave{c_{k\sigma}(\tau)c_{k\sigma}\adj}_0
    = n_{k\sigma}^l f(-\xi_k^l) e^{-\tau\xi_k^l} + (l \to u)
\end{equation}
for $\xi_k^l = \xi_k$ and $\xi_k^u = \xi_k+U$, $n_{k\sigma}^u = \ave{n_{k\bar{\sigma}}}_0$ and $n_{k\sigma}^l = 1-n_{k\sigma}^u$, and $f(\omega)$ the Fermi function at temperature $T$, we have
\begin{align}
    \chi_0(i\nu_n) &= \chi_0^{l l} + \chi_0^{u u} + \chi_0^{l u} + \chi_0^{u l} \\
    \chi_0^{a b} &= \frac{1}{L^d} \sum_k n_{k\up}^a n_{-k\dn}^b \frac{f(\omega_{k}^a)+f(\omega_{-k}^b)-1}{i\nu_n-\omega_{k}^a-\omega_{-k}^b}
\end{align}

where the superscripts $a b$ may represent $l l$, $u u$, $l u$, or $u l$.
Because of the factors $f(\omega_{k}^l)+f(\omega_{-k}^u)-1$, the cross terms $\chi_0^{l u}$ and $\chi_0^{u l}$ (between the lower and upper Hubbard bands) contribute no low-energy spectral weight when $T \ll U$ and are dropped hereafter.
Using $\xi_k = \xi_{-k}$ and $\ave{n_{k\up}} = \ave{n_{k\dn}}$ and changing variables, we finally arrive at
\begin{align}
\chi_0(0) &= \int d\omega \; N'(\omega) \frac{\tanh{\frac{\beta\omega}{2}}}{2 \omega} \\
 L^d N'(\omega) &= \sum_{k \in \Omega_0} \delta(\omega-\xi_k^l) + \sum_{k \in \Omega_2} \delta(\omega-\xi_k^u) \\
&+ \frac{1}{4}\sum_{k \in \Omega_1} \delta(\omega-\xi_k^l) + \delta(\omega-\xi_k^u).
\end{align}
Here, $N'(\omega)$ is an effective density of states, similar to but not equal to the HK model's single-particle density of states $N(\omega) = \frac{1}{L^d} \sum_k \frac{-1}{\pi} \Im G(k,\omega+i0^+)$, which has a factor of $\frac{1}{2}$ rather than $\frac{1}{4}$ before the sum over the singly occupied region. By the same manipulations as in the free fermion case, it is clear that $\chi_0(0)$ grows as $\ln\frac{1}{T}$ at low temperature. Hence with any nonzero pairing strength $g$, $\chi(0)$ diverges at the transition temperature $T_c \propto e^{-1/(N'(0) g)}$ (see supplementary text \cite{supp}). In Fig.~\ref{fig:EbTc}, the transition temperatures are calculated explicitly for a variety of parameters.

From both the Cooper argument and a direct calculation of the pair susceptibility, we have established that the HK model has a superconducting instability. We now seek to characterize the ground state of the model in the presence of a nonzero pairing interaction. We work with the variational wavefunction
\begin{equation}
\ket{\psi} = \prod_{k > 0} \qty(x_k + y_k b_k^\dagger b_{-k}^\dagger + \frac{z_k}{\sqrt{2}} \qty(b_k^\dagger + b_{-k}^\dagger)) \ket{0}
\end{equation}
normalized by $\abs{x_k}^2 + \abs{y_k}^2 + \abs{z_k}^2 = 1$. This is a generalization of the BCS wavefunction, which corresponds to $x_k = u_k^2$, $y_k = v_k^2$, $z_k = \sqrt{2} u_k v_k$. The utility of this generalized wavefunction is that the state with $x_k = 1$ for $k \in \Omega_0$, $y_k = 1$ for $k \in \Omega_2$, and $z_k = 1$ for $k \in \Omega_1$, is a ground state of the HK model. Minimizing the energy variationally leads to two equations
\begin{align}
\xi_k^l x_k z_k &= \qty(x_k^2 - z_k^2) \frac{g}{L^d} \sum_{p>0} \qty(x_p z_p + z_p y_p) \\
\xi_k^u y_k z_k &= \qty(z_k^2 - y_k^2) \frac{g}{L^d} \sum_{p>0} \qty(x_p z_p + z_p y_p),
\end{align}
after taking the limit $g \ll U, W$. Details are provided in the supplementary text \cite{supp}. After a few changes of variables, we obtain a gap equation
\begin{align}
1 &= \frac{g}{2} \int \dd{\omega} \frac{N''(\omega)}{\sqrt{\omega^2 + \Delta^2}} \\
N''(\omega) &= \sum_{k \in \Omega_0} \delta(\omega-\xi_k^l) + \sum_{k \in \Omega_2} \delta(\omega-\xi_k^u) \nonumber\\
&+ \sum_{k \in \Omega_1} \delta(\omega-\xi_k^l) + \delta(\omega-\xi_k^u)\nonumber.
\label{gapeq}
\end{align}
This is the BCS gap equation except for the effective density of states $N''(\omega)$. The solution has $\Delta \propto e^{-1/(N''(0) g)}$, which is verified in the suppementary text \cite{supp} for a specific example. Note that $N''(\omega)$ is different from $N'(\omega)$ which controls $T_c$, as there is no factor of $\frac{1}{4}$ before the sum $\sum_{k \in \Omega_1}$. Because this sum over the singly occupied region affects the low energy spectra, $N''(0) > N'(0)$ and the superconducting gap-to-transition temperature ratio diverges as $e^{3/(10 \rho(\mu) g)}$ for $g \to 0$.
This is in contrast to the universal BCS result $\frac{2\Delta}{T_c} = 3.53\dots$ for $s$-wave pairing in the weak coupling limit. Therefore, despite apparent mathematical similarities between pairing in the HK model and BCS pairing of free fermions, the presence of the singly occupied region $\Omega_1$ leads to qualitatively different phenomena.

\begin{figure}
\centering
\includegraphics{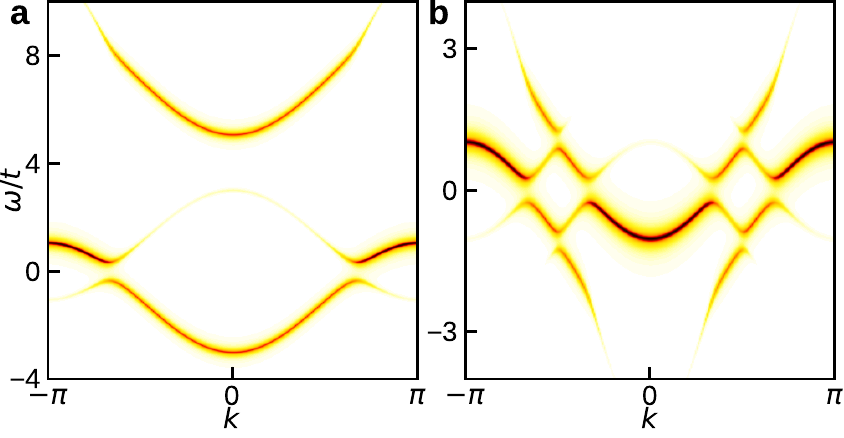}
\includegraphics{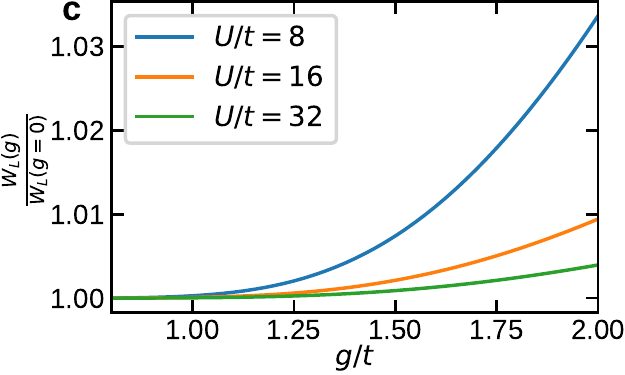}
\caption{
{\bf Spectral function in the superconducting ground state.} {\bf a}, Doped Mott state. Parameters are $U/t = 8$, $g/t=1.75$, $\mu/t=1$. {\bf b}, Half-filled metal. Parameters are $U/t = 2$, $g/t=1$, $\mu/t=1$. Darker colors indicate higher intensity. Details of the calculation are provided in the Supplementary Materials\cite{supp}. {\bf c}, Integrated spectral weight in the lower Hubbard band ($-4 < \omega/t < 4$), at chemical potential $\mu/t=1$, relative to the value for $g=0$. The increase here is consistent with the experimental trends\cite{marel1}.
}
\label{fig:Akw}
\end{figure}

The elementary excitations of a BCS superconductor are Bogoliubov quasiparticles  $\gamma_{k \sigma} = u_k c_{k \sigma} - \sigma v_k c_{-k \overline{\sigma}}^\dagger$. The excitations of the superconducting state of the HK model with pairing cannot be the same. This is for the same reason that $c_{k\sigma}^\dagger$ is not an elementary excitation of the HK model, namely that in the singly occupied region, both the upper and lower Hubbard bands have nonzero spectral weight. On the other hand, the Green functions for the holon and doublon excitations, $\zeta_{k\sigma}$ and $\eta_{k\sigma}$, have weight only in the lower and upper Hubbard bands, respectively; these composite operators describe the elementary excitations of the HK model. Upon turning on pairing and entering the superconducting state, the new excitations are given by mixing of the composite operators and their conjugates
\begin{align}
\gamma_{k \sigma}^{l} &\propto \sqrt{2} x_k \zeta_{k \sigma}^\dagger - \sigma z_k \zeta_{-k \overline{\sigma}} \\
\gamma_{k \sigma}^{u} &\propto z_k \eta_{k \sigma}^\dagger - \sigma \sqrt{2} y_k \eta_{-k \overline{\sigma}}.
\end{align}
It is straightforward to check that
\begin{align}
\gamma_{k\sigma}^{u/l} \ket{\psi} &= 0 \\
\ev{\gamma_{k\sigma}^{u/l} H \qty(\gamma_{k\sigma}^{u/l})^\dagger}{\psi} &= \ev{H}{\psi} + E_k^{u/l},
\end{align}
where $E_k^{u/l} = \sqrt{{\xi_k^{u/l}}^2 + \Delta^2}$. Therefore, $\gamma_{k\sigma}^{u/l}$ are analogous to the Bogoliubov quasiparticle excitations of a BCS superconductor but composed of doublons or holons, indicating that the non-Fermi liquid nature of the metallic state carries over into the superconducting state.
To further reveal the role of such excitations, we compute the spectral function in the superconducting state, treating the pairing interaction ($g$) at the mean-field level and the Mott interaction ($U$) exactly.  Plots for the doped Mott insulator and the half-filled metal are provided in Fig.~\ref{fig:Akw}a and b, respectively. Here, we see that these excitations are present in the single-particle spectral function at energies $\pm E_k^{u/l}$. Where the Hubbard bands cross the chemical potential ($\omega/t=0$), we observe the superconducting gap and associated back-bending of the dispersions. Therefore despite the unconventional nature of the excitations of the superconducting state as discussed previously, their spectroscopic signatures appear BCS-like.

Because the quasiparticles in the superconducting state involve the composite Hubbard operators, the HK model has the ingredients to explain the experimentally observed\cite{marel1} color change.  Indeed it does as Fig.~\ref{fig:Akw}c attests.   Namely, the integrated weight of the lower band in the superconducting state relative to the weight at  $g=0$ is a monotonically increasing function of the pair coupling constant, $g$.  The size of this increase is consistent with the 6\% increase seen experimentally\cite{marel1}.  The origin of this effect in the HK model is simple.  The pairing interaction ruins the commutativity of the kinetic  and potential energy terms in Eq. (\ref{Htot}).  This will result in dynamical mixing between the upper and lower Hubbard bands which is well documented\cite{sawatzky,eskes,RMP} to increase the spectral weight in the lower band as it increases the number of low-energy degrees of freedom.  Conservation of the spectral weight necessitates an equivalent decrease in the high-energy part of the spectrum as seen experimentally\cite{marel1}.  Since it is the pairing term which leads to a breakdown of the integrability of the HK model,  DSWT in the HK model is a direct result of opening the superconducting gap.   As the subsequent curves show in Fig.\ref{fig:Akw}c, the integrated weight falls off as $1/U$ as is expected for dynamical mixing across the Hubbard bands.  Such mixing is a general consequence of Mott physics and hence transcends the HK model.  Such dynamical mixing can certainly be enhanced by instantaneous pairing interactions\cite{scalapino} as such interactions involve all energy scales.

\begin{figure}
\centering
\includegraphics{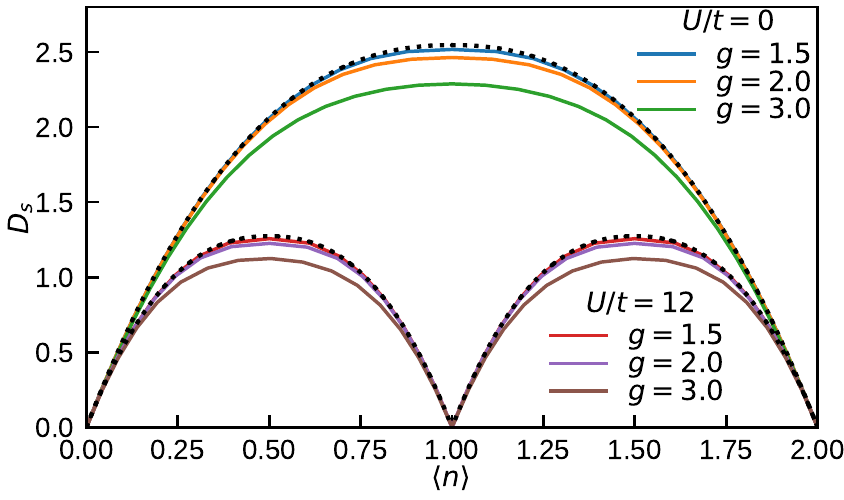}
\caption{
{\bf Ground state superfluid stiffness.} Solid lines are the superfluid stiffness for various values of interaction strength $U/t$ and pairing strength $g/t$. Dotted lines are proportional to the kinetic energy when $g = 0$, representing an upper bound on the superfluid stiffness.
}
\label{fig:stiff}
\end{figure}

The consequences of Mottness on the superconducting state extend also to the superfluid stiffness. Again treating the pairing interaction at the mean-field level, we have calculated the superfluid stiffness in the superconducting ground state in Fig.~\ref{fig:stiff}.  As compared to a regular BCS superconductor, the superfluid stiffness in the HK model is significantly suppressed, particularly near half-filling. This is a direct consequence of the Mott interaction: proximity to the Mott insulator reduces the kinetic energy and hence the effective carrier density, which is an upper bound to the superfluid stiffness.  While models with disorder and  fluctuations of the pair-field also lead to a suppression, Mottness supercedes all such effects and hence the effect we have found here is quite general.

Given the recent spate of papers on superconductivity in the absence of quasiparticles\cite{patel2018,cai2018,wang2019,esterlis2019,chowdhury2019,hauck2019,wu2019,chubukov2019}, our approach offers a systematic Hamiltonian-based approach to the breakdown of the quasiparticle picture without invoking randomness.  As remarked previously, recent work\cite{chandan} on the Luttinger surface applies strictly to the insulator where the susceptibility appears to diverge, exhibiting SYK dynamics. Our analysis has revealed here that when both upper and lower Hubbard bands carry spectral weight, the essence of Mottness, the fundamental excitations of either the metallic or superconducting states of a doped Mott insulator cannot be described by conventional quasiparticles.

This is a conclusion that applies also to the cuprates, in which the spectral weight of the upper Hubbard band in hole-doped compounds has been observed and compared to calculations of the Hubbard model\cite{RMP,yang2017}. Although we do not know precisely the excitations of the strange metal normal state of cuprates\cite{zaanen2019}, there is overwhelming evidence that they are not Fermi liquid quasiparticles.  These arguments extend this notion to the superconducting state. Our findings for the HK model and the origin of the color change which should hold for the Hubbard model, thus challenge the assumption that the appearance of coherent peaks and back-bending in the spectral function of superconducting cuprates is a signature of regular Bogoliubov quasiparticles. Conversely, they suggest that detailed studies of the superconducting state and its excitations can help unravel the mysteries of the normal state as a general consequence of UV-IR mixing\cite{Zaanen}.

While the HK model is complex enough to capture zeros of the Green function and their associated consequences on the metallic state and on the superconducting instability, it does not support dynamical spectral weight transfer unless pairing is included.
A promising line of inquiry would be to see how stable the present results are to such dynamical mixing arising from repulsive interactions.
Whether a renormalization principle can be established to show that the excitations on a zero surface are impervious to such mixing remains an open question.


\begin{thebibliography}{36}%
\makeatletter
\providecommand \@ifxundefined [1]{%
 \@ifx{#1\undefined}
}%
\providecommand \@ifnum [1]{%
 \ifnum #1\expandafter \@firstoftwo
 \else \expandafter \@secondoftwo
 \fi
}%
\providecommand \@ifx [1]{%
 \ifx #1\expandafter \@firstoftwo
 \else \expandafter \@secondoftwo
 \fi
}%
\providecommand \natexlab [1]{#1}%
\providecommand \enquote  [1]{``#1''}%
\providecommand \bibnamefont  [1]{#1}%
\providecommand \bibfnamefont [1]{#1}%
\providecommand \citenamefont [1]{#1}%
\providecommand \href@noop [0]{\@secondoftwo}%
\providecommand \href [0]{\begingroup \@sanitize@url \@href}%
\providecommand \@href[1]{\@@startlink{#1}\@@href}%
\providecommand \@@href[1]{\endgroup#1\@@endlink}%
\providecommand \@sanitize@url [0]{\catcode `\\12\catcode `\$12\catcode
  `\&12\catcode `\#12\catcode `\^12\catcode `\_12\catcode `\%12\relax}%
\providecommand \@@startlink[1]{}%
\providecommand \@@endlink[0]{}%
\providecommand \url  [0]{\begingroup\@sanitize@url \@url }%
\providecommand \@url [1]{\endgroup\@href {#1}{\urlprefix }}%
\providecommand \urlprefix  [0]{URL }%
\providecommand \Eprint [0]{\href }%
\providecommand \doibase [0]{http://dx.doi.org/}%
\providecommand \selectlanguage [0]{\@gobble}%
\providecommand \bibinfo  [0]{\@secondoftwo}%
\providecommand \bibfield  [0]{\@secondoftwo}%
\providecommand \translation [1]{[#1]}%
\providecommand \BibitemOpen [0]{}%
\providecommand \bibitemStop [0]{}%
\providecommand \bibitemNoStop [0]{.\EOS\space}%
\providecommand \EOS [0]{\spacefactor3000\relax}%
\providecommand \BibitemShut  [1]{\csname bibitem#1\endcsname}%
\let\auto@bib@innerbib\@empty
\bibitem [{\citenamefont {Hatsugai}\ and\ \citenamefont
  {Kohmoto}(1992)}]{hk1992}%
  \BibitemOpen
  \bibfield  {author} {\bibinfo {author} {\bibfnamefont {Y.}~\bibnamefont
  {Hatsugai}}\ and\ \bibinfo {author} {\bibfnamefont {M.}~\bibnamefont
  {Kohmoto}},\ }\href {\doibase 10.1143/JPSJ.61.2056} {\bibfield  {journal}
  {\bibinfo  {journal} {Journal of the Physical Society of Japan}\ }\textbf
  {\bibinfo {volume} {61}},\ \bibinfo {pages} {2056} (\bibinfo {year}
  {1992})}\BibitemShut {NoStop}%
\bibitem [{\citenamefont {Baskaran}(1991)}]{baskaran1991exactly}%
  \BibitemOpen
  \bibfield  {author} {\bibinfo {author} {\bibfnamefont {G.}~\bibnamefont
  {Baskaran}},\ }\href@noop {} {\bibfield  {journal} {\bibinfo  {journal}
  {Modern Physics Letters B}\ }\textbf {\bibinfo {volume} {5}},\ \bibinfo
  {pages} {643} (\bibinfo {year} {1991})}\BibitemShut {NoStop}%
\bibitem [{\citenamefont {Molegraaf}\ \emph {et~al.}(2002)\citenamefont
  {Molegraaf}, \citenamefont {Presura}, \citenamefont {van~der Marel},
  \citenamefont {Kes},\ and\ \citenamefont {Li}}]{marel1}%
  \BibitemOpen
  \bibfield  {author} {\bibinfo {author} {\bibfnamefont {H.~J.~A.}\
  \bibnamefont {Molegraaf}}, \bibinfo {author} {\bibfnamefont {C.}~\bibnamefont
  {Presura}}, \bibinfo {author} {\bibfnamefont {D.}~\bibnamefont {van~der
  Marel}}, \bibinfo {author} {\bibfnamefont {P.~H.}\ \bibnamefont {Kes}}, \
  and\ \bibinfo {author} {\bibfnamefont {M.}~\bibnamefont {Li}},\ }\href
  {http://www.sciencemag.org/cgi/content/abstract/295/5563/2239} {\bibfield
  {journal} {\bibinfo  {journal} {Science}\ }\textbf {\bibinfo {volume}
  {295}},\ \bibinfo {pages} {2239} (\bibinfo {year} {2002})}\BibitemShut
  {NoStop}%
\bibitem [{\citenamefont {Cooper}(1956)}]{cooper}%
  \BibitemOpen
  \bibfield  {author} {\bibinfo {author} {\bibfnamefont {L.~N.}\ \bibnamefont
  {Cooper}},\ }\href {\doibase 10.1103/PhysRev.104.1189} {\bibfield  {journal}
  {\bibinfo  {journal} {Phys. Rev.}\ }\textbf {\bibinfo {volume} {104}},\
  \bibinfo {pages} {1189} (\bibinfo {year} {1956})}\BibitemShut {NoStop}%
\bibitem [{\citenamefont {Shankar}(1994)}]{shankar}%
  \BibitemOpen
  \bibfield  {author} {\bibinfo {author} {\bibfnamefont {R.}~\bibnamefont
  {Shankar}},\ }\href {\doibase 10.1103/RevModPhys.66.129} {\bibfield
  {journal} {\bibinfo  {journal} {Rev. Mod. Phys.}\ }\textbf {\bibinfo {volume}
  {66}},\ \bibinfo {pages} {129} (\bibinfo {year} {1994})}\BibitemShut
  {NoStop}%
\bibitem [{\citenamefont {Polchinski}(1992)}]{polchinski}%
  \BibitemOpen
  \bibfield  {author} {\bibinfo {author} {\bibfnamefont {J.}~\bibnamefont
  {Polchinski}},\ }\href@noop {} {\bibfield  {journal} {\bibinfo  {journal}
  {arXiv:hep-th/9210046v2}\ } (\bibinfo {year} {1992})}\BibitemShut {NoStop}%
\bibitem [{\citenamefont {Benfatto}\ and\ \citenamefont
  {Gallavotti}(1990)}]{benfatto}%
  \BibitemOpen
  \bibfield  {author} {\bibinfo {author} {\bibfnamefont {G.}~\bibnamefont
  {Benfatto}}\ and\ \bibinfo {author} {\bibfnamefont {G.}~\bibnamefont
  {Gallavotti}},\ }\href {\doibase 10.1103/PhysRevB.42.9967} {\bibfield
  {journal} {\bibinfo  {journal} {Phys. Rev. B}\ }\textbf {\bibinfo {volume}
  {42}},\ \bibinfo {pages} {9967} (\bibinfo {year} {1990})}\BibitemShut
  {NoStop}%
\bibitem [{\citenamefont {{Phillips}}(2010)}]{RMP}%
  \BibitemOpen
  \bibfield  {author} {\bibinfo {author} {\bibfnamefont {P.}~\bibnamefont
  {{Phillips}}},\ }\href {\doibase 10.1103/RevModPhys.82.1719} {\bibfield
  {journal} {\bibinfo  {journal} {Reviews of Modern Physics}\ }\textbf
  {\bibinfo {volume} {82}},\ \bibinfo {pages} {1719} (\bibinfo {year}
  {2010})},\ \Eprint {http://arxiv.org/abs/1001.5270} {arXiv:1001.5270
  [cond-mat.str-el]} \BibitemShut {NoStop}%
\bibitem [{\citenamefont {Dzyaloshinskii}(2003)}]{dzy}%
  \BibitemOpen
  \bibfield  {author} {\bibinfo {author} {\bibfnamefont {I.}~\bibnamefont
  {Dzyaloshinskii}},\ }\href {\doibase 10.1103/PhysRevB.68.085113} {\bibfield
  {journal} {\bibinfo  {journal} {Phys. Rev. B}\ }\textbf {\bibinfo {volume}
  {68}},\ \bibinfo {pages} {085113} (\bibinfo {year} {2003})}\BibitemShut
  {NoStop}%
\bibitem [{\citenamefont {Meinders}\ \emph {et~al.}(1993)\citenamefont
  {Meinders}, \citenamefont {Eskes},\ and\ \citenamefont
  {Sawatzky}}]{sawatzky}%
  \BibitemOpen
  \bibfield  {author} {\bibinfo {author} {\bibfnamefont {M.~B.~J.}\
  \bibnamefont {Meinders}}, \bibinfo {author} {\bibfnamefont {H.}~\bibnamefont
  {Eskes}}, \ and\ \bibinfo {author} {\bibfnamefont {G.~A.}\ \bibnamefont
  {Sawatzky}},\ }\href {\doibase 10.1103/PhysRevB.48.3916} {\bibfield
  {journal} {\bibinfo  {journal} {Phys. Rev. B}\ }\textbf {\bibinfo {volume}
  {48}},\ \bibinfo {pages} {3916} (\bibinfo {year} {1993})}\BibitemShut
  {NoStop}%
\bibitem [{\citenamefont {Eskes}\ \emph {et~al.}(1994)\citenamefont {Eskes},
  \citenamefont {Ole\ifmmode~\acute{s}\else \'{s}\fi{}}, \citenamefont
  {Meinders},\ and\ \citenamefont {Stephan}}]{eskes}%
  \BibitemOpen
  \bibfield  {author} {\bibinfo {author} {\bibfnamefont {H.}~\bibnamefont
  {Eskes}}, \bibinfo {author} {\bibfnamefont {A.~M.}\ \bibnamefont
  {Ole\ifmmode~\acute{s}\else \'{s}\fi{}}}, \bibinfo {author} {\bibfnamefont
  {M.~B.~J.}\ \bibnamefont {Meinders}}, \ and\ \bibinfo {author} {\bibfnamefont
  {W.}~\bibnamefont {Stephan}},\ }\href {\doibase 10.1103/PhysRevB.50.17980}
  {\bibfield  {journal} {\bibinfo  {journal} {Phys. Rev. B}\ }\textbf {\bibinfo
  {volume} {50}},\ \bibinfo {pages} {17980} (\bibinfo {year}
  {1994})}\BibitemShut {NoStop}%
\bibitem [{\citenamefont {Chen}\ \emph {et~al.}(1991)\citenamefont {Chen},
  \citenamefont {Sette}, \citenamefont {Ma}, \citenamefont {Hybertsen},
  \citenamefont {Stechel}, \citenamefont {Foulkes}, \citenamefont {Schulter},
  \citenamefont {Cheong}, \citenamefont {Cooper}, \citenamefont {Rupp},
  \citenamefont {Batlogg}, \citenamefont {Soo}, \citenamefont {Ming},
  \citenamefont {Krol},\ and\ \citenamefont {Kao}}]{chen}%
  \BibitemOpen
  \bibfield  {author} {\bibinfo {author} {\bibfnamefont {C.~T.}\ \bibnamefont
  {Chen}}, \bibinfo {author} {\bibfnamefont {F.}~\bibnamefont {Sette}},
  \bibinfo {author} {\bibfnamefont {Y.}~\bibnamefont {Ma}}, \bibinfo {author}
  {\bibfnamefont {M.~S.}\ \bibnamefont {Hybertsen}}, \bibinfo {author}
  {\bibfnamefont {E.~B.}\ \bibnamefont {Stechel}}, \bibinfo {author}
  {\bibfnamefont {W.~M.~C.}\ \bibnamefont {Foulkes}}, \bibinfo {author}
  {\bibfnamefont {M.}~\bibnamefont {Schulter}}, \bibinfo {author}
  {\bibfnamefont {S.-W.}\ \bibnamefont {Cheong}}, \bibinfo {author}
  {\bibfnamefont {A.~S.}\ \bibnamefont {Cooper}}, \bibinfo {author}
  {\bibfnamefont {L.~W.}\ \bibnamefont {Rupp}}, \bibinfo {author}
  {\bibfnamefont {B.}~\bibnamefont {Batlogg}}, \bibinfo {author} {\bibfnamefont
  {Y.~L.}\ \bibnamefont {Soo}}, \bibinfo {author} {\bibfnamefont {Z.~H.}\
  \bibnamefont {Ming}}, \bibinfo {author} {\bibfnamefont {A.}~\bibnamefont
  {Krol}}, \ and\ \bibinfo {author} {\bibfnamefont {Y.~H.}\ \bibnamefont
  {Kao}},\ }\href {\doibase 10.1103/PhysRevLett.66.104} {\bibfield  {journal}
  {\bibinfo  {journal} {Phys. Rev. Lett.}\ }\textbf {\bibinfo {volume} {66}},\
  \bibinfo {pages} {104} (\bibinfo {year} {1991})}\BibitemShut {NoStop}%
\bibitem [{\citenamefont {Uchida}\ \emph {et~al.}(1991)\citenamefont {Uchida},
  \citenamefont {Ido}, \citenamefont {Takagi}, \citenamefont {Arima},
  \citenamefont {Tokura},\ and\ \citenamefont {Tajima}}]{uchida}%
  \BibitemOpen
  \bibfield  {author} {\bibinfo {author} {\bibfnamefont {S.}~\bibnamefont
  {Uchida}}, \bibinfo {author} {\bibfnamefont {T.}~\bibnamefont {Ido}},
  \bibinfo {author} {\bibfnamefont {H.}~\bibnamefont {Takagi}}, \bibinfo
  {author} {\bibfnamefont {T.}~\bibnamefont {Arima}}, \bibinfo {author}
  {\bibfnamefont {Y.}~\bibnamefont {Tokura}}, \ and\ \bibinfo {author}
  {\bibfnamefont {S.}~\bibnamefont {Tajima}},\ }\href {\doibase
  10.1103/PhysRevB.43.7942} {\bibfield  {journal} {\bibinfo  {journal} {Phys.
  Rev. B}\ }\textbf {\bibinfo {volume} {43}},\ \bibinfo {pages} {7942}
  (\bibinfo {year} {1991})}\BibitemShut {NoStop}%
\bibitem [{\citenamefont {Santander-Syro}\ \emph {et~al.}(2003)\citenamefont
  {Santander-Syro}, \citenamefont {Lobo}, \citenamefont {Bontemps},
  \citenamefont {Konstantinovic}, \citenamefont {Li},\ and\ \citenamefont
  {Raffy}}]{bontemps}%
  \BibitemOpen
  \bibfield  {author} {\bibinfo {author} {\bibfnamefont {A.~F.}\ \bibnamefont
  {Santander-Syro}}, \bibinfo {author} {\bibfnamefont {R.~P. S.~M.}\
  \bibnamefont {Lobo}}, \bibinfo {author} {\bibfnamefont {N.}~\bibnamefont
  {Bontemps}}, \bibinfo {author} {\bibfnamefont {Z.}~\bibnamefont
  {Konstantinovic}}, \bibinfo {author} {\bibfnamefont {Z.~Z.}\ \bibnamefont
  {Li}}, \ and\ \bibinfo {author} {\bibfnamefont {H.}~\bibnamefont {Raffy}},\
  }\href {http://stacks.iop.org/0295-5075/62/568} {\bibfield  {journal}
  {\bibinfo  {journal} {EPL (Europhysics Letters)}\ }\textbf {\bibinfo {volume}
  {62}},\ \bibinfo {pages} {568} (\bibinfo {year} {2003})}\BibitemShut
  {NoStop}%
\bibitem [{\citenamefont {{Setty}}(2019)}]{chandan}%
  \BibitemOpen
  \bibfield  {author} {\bibinfo {author} {\bibfnamefont {C.}~\bibnamefont
  {{Setty}}},\ }\href@noop {} {\bibfield  {journal} {\bibinfo  {journal} {arXiv
  e-prints}\ ,\ \bibinfo {eid} {arXiv:1908.09831}} (\bibinfo {year} {2019})},\
  \Eprint {http://arxiv.org/abs/1908.09831} {arXiv:1908.09831
  [cond-mat.str-el]} \BibitemShut {NoStop}%
\bibitem [{\citenamefont {Bo{\v z}ovi{\'c}}\ \emph {et~al.}(2016)\citenamefont
  {Bo{\v z}ovi{\'c}}, \citenamefont {He}, \citenamefont {Wu},\ and\
  \citenamefont {Bollinger}}]{bozovic}%
  \BibitemOpen
  \bibfield  {author} {\bibinfo {author} {\bibfnamefont {I.}~\bibnamefont
  {Bo{\v z}ovi{\'c}}}, \bibinfo {author} {\bibfnamefont {X.}~\bibnamefont
  {He}}, \bibinfo {author} {\bibfnamefont {J.}~\bibnamefont {Wu}}, \ and\
  \bibinfo {author} {\bibfnamefont {A.~T.}\ \bibnamefont {Bollinger}},\ }\href
  {\doibase 10.1038/nature19061} {\bibfield  {journal} {\bibinfo  {journal}
  {Nature}\ }\textbf {\bibinfo {volume} {536}},\ \bibinfo {pages} {309}
  (\bibinfo {year} {2016})}\BibitemShut {NoStop}%
\bibitem [{\citenamefont {Sachdev}\ and\ \citenamefont {Ye}(1993)}]{sy1993}%
  \BibitemOpen
  \bibfield  {author} {\bibinfo {author} {\bibfnamefont {S.}~\bibnamefont
  {Sachdev}}\ and\ \bibinfo {author} {\bibfnamefont {J.}~\bibnamefont {Ye}},\
  }\href {\doibase 10.1103/PhysRevLett.70.3339} {\bibfield  {journal} {\bibinfo
   {journal} {Phys. Rev. Lett.}\ }\textbf {\bibinfo {volume} {70}},\ \bibinfo
  {pages} {3339} (\bibinfo {year} {1993})}\BibitemShut {NoStop}%
\bibitem [{\citenamefont {Kitaev}(2015)}]{k2015}%
  \BibitemOpen
  \bibfield  {author} {\bibinfo {author} {\bibfnamefont {A.}~\bibnamefont
  {Kitaev}},\ }\href@noop {} {\bibfield  {journal} {\bibinfo  {journal}
  {http://online.kitp.ucsb.edu/online/entangled15/kitaev/}\ } (\bibinfo {year}
  {2015})}\BibitemShut {NoStop}%
\bibitem [{\citenamefont {Dave}\ \emph {et~al.}(2013)\citenamefont {Dave},
  \citenamefont {Phillips},\ and\ \citenamefont {Kane}}]{dave2013}%
  \BibitemOpen
  \bibfield  {author} {\bibinfo {author} {\bibfnamefont {K.~B.}\ \bibnamefont
  {Dave}}, \bibinfo {author} {\bibfnamefont {P.~W.}\ \bibnamefont {Phillips}},
  \ and\ \bibinfo {author} {\bibfnamefont {C.~L.}\ \bibnamefont {Kane}},\
  }\href {\doibase 10.1103/PhysRevLett.110.090403} {\bibfield  {journal}
  {\bibinfo  {journal} {Physical Review Letters}\ }\textbf {\bibinfo {volume}
  {110}} (\bibinfo {year} {2013}),\ 10.1103/PhysRevLett.110.090403},\ \Eprint
  {http://arxiv.org/abs/1207.4201} {arXiv:1207.4201} \BibitemShut {NoStop}%
\bibitem [{sup()}]{supp}%
  \BibitemOpen
  \href@noop {} {}\bibinfo {note} {See supplementary materials}\BibitemShut
  {NoStop}%
\bibitem [{\citenamefont {MUTHUKUMAR}\ and\ \citenamefont
  {BASKARAN}(1994)}]{bask}%
  \BibitemOpen
  \bibfield  {author} {\bibinfo {author} {\bibfnamefont {V.~N.}\ \bibnamefont
  {MUTHUKUMAR}}\ and\ \bibinfo {author} {\bibfnamefont {G.}~\bibnamefont
  {BASKARAN}},\ }\href {\doibase 10.1142/S0217984994000716} {\bibfield
  {journal} {\bibinfo  {journal} {Modern Physics Letters B}\ }\textbf {\bibinfo
  {volume} {08}},\ \bibinfo {pages} {699} (\bibinfo {year} {1994})},\ \Eprint
  {http://arxiv.org/abs/https://doi.org/10.1142/S0217984994000716}
  {https://doi.org/10.1142/S0217984994000716} \BibitemShut {NoStop}%
\bibitem [{\citenamefont {Metlitski}\ \emph {et~al.}(2015)\citenamefont
  {Metlitski}, \citenamefont {Mross}, \citenamefont {Sachdev},\ and\
  \citenamefont {Senthil}}]{mross}%
  \BibitemOpen
  \bibfield  {author} {\bibinfo {author} {\bibfnamefont {M.~A.}\ \bibnamefont
  {Metlitski}}, \bibinfo {author} {\bibfnamefont {D.~F.}\ \bibnamefont
  {Mross}}, \bibinfo {author} {\bibfnamefont {S.}~\bibnamefont {Sachdev}}, \
  and\ \bibinfo {author} {\bibfnamefont {T.}~\bibnamefont {Senthil}},\ }\href
  {\doibase 10.1103/PhysRevB.91.115111} {\bibfield  {journal} {\bibinfo
  {journal} {Phys. Rev. B}\ }\textbf {\bibinfo {volume} {91}},\ \bibinfo
  {pages} {115111} (\bibinfo {year} {2015})}\BibitemShut {NoStop}%
\bibitem [{\citenamefont {{Varlamov}}\ \emph {et~al.}(2018)\citenamefont
  {{Varlamov}}, \citenamefont {{Galda}},\ and\ \citenamefont
  {{Glatz}}}]{varlamov-galda-glatz}%
  \BibitemOpen
  \bibfield  {author} {\bibinfo {author} {\bibfnamefont {A.~A.}\ \bibnamefont
  {{Varlamov}}}, \bibinfo {author} {\bibfnamefont {A.}~\bibnamefont {{Galda}}},
  \ and\ \bibinfo {author} {\bibfnamefont {A.}~\bibnamefont {{Glatz}}},\ }\href
  {\doibase 10.1103/RevModPhys.90.015009} {\bibfield  {journal} {\bibinfo
  {journal} {Reviews of Modern Physics}\ }\textbf {\bibinfo {volume} {90}},\
  \bibinfo {eid} {015009} (\bibinfo {year} {2018})}\BibitemShut {NoStop}%
\bibitem [{\citenamefont {Maier}\ \emph {et~al.}(2008)\citenamefont {Maier},
  \citenamefont {Poilblanc},\ and\ \citenamefont {Scalapino}}]{scalapino}%
  \BibitemOpen
  \bibfield  {author} {\bibinfo {author} {\bibfnamefont {T.~A.}\ \bibnamefont
  {Maier}}, \bibinfo {author} {\bibfnamefont {D.}~\bibnamefont {Poilblanc}}, \
  and\ \bibinfo {author} {\bibfnamefont {D.~J.}\ \bibnamefont {Scalapino}},\
  }\href {\doibase 10.1103/PhysRevLett.100.237001} {\bibfield  {journal}
  {\bibinfo  {journal} {Phys. Rev. Lett.}\ }\textbf {\bibinfo {volume} {100}},\
  \bibinfo {pages} {237001} (\bibinfo {year} {2008})}\BibitemShut {NoStop}%
\bibitem [{\citenamefont {Patel}\ \emph {et~al.}(2018)\citenamefont {Patel},
  \citenamefont {Lawler},\ and\ \citenamefont {Kim}}]{patel2018}%
  \BibitemOpen
  \bibfield  {author} {\bibinfo {author} {\bibfnamefont {A.~A.}\ \bibnamefont
  {Patel}}, \bibinfo {author} {\bibfnamefont {M.~J.}\ \bibnamefont {Lawler}}, \
  and\ \bibinfo {author} {\bibfnamefont {E.-A.}\ \bibnamefont {Kim}},\ }\href
  {\doibase 10.1103/PhysRevLett.121.187001} {\bibfield  {journal} {\bibinfo
  {journal} {Phys. Rev. Lett.}\ }\textbf {\bibinfo {volume} {121}},\ \bibinfo
  {pages} {187001} (\bibinfo {year} {2018})}\BibitemShut {NoStop}%
\bibitem [{\citenamefont {Cai}\ and\ \citenamefont {Ge}(2018)}]{cai2018}%
  \BibitemOpen
  \bibfield  {author} {\bibinfo {author} {\bibfnamefont {W.}~\bibnamefont
  {Cai}}\ and\ \bibinfo {author} {\bibfnamefont {X.-H.}\ \bibnamefont {Ge}},\
  }\href {https://arxiv.org/abs/1809.01846} {\bibfield  {journal} {\bibinfo
  {journal} {arXiv:1809.01846}\ } (\bibinfo {year} {2018})}\BibitemShut
  {NoStop}%
\bibitem [{\citenamefont {Wang}(2019)}]{wang2019}%
  \BibitemOpen
  \bibfield  {author} {\bibinfo {author} {\bibfnamefont {Y.}~\bibnamefont
  {Wang}},\ }\href {https://arxiv.org/abs/1904.07240} {\bibfield  {journal}
  {\bibinfo  {journal} {arXiv:1904.07240}\ } (\bibinfo {year}
  {2019})}\BibitemShut {NoStop}%
\bibitem [{\citenamefont {Esterlis}\ and\ \citenamefont
  {Schmalian}(2019)}]{esterlis2019}%
  \BibitemOpen
  \bibfield  {author} {\bibinfo {author} {\bibfnamefont {I.}~\bibnamefont
  {Esterlis}}\ and\ \bibinfo {author} {\bibfnamefont {J.}~\bibnamefont
  {Schmalian}},\ }\href {\doibase 10.1103/PhysRevB.100.115132} {\bibfield
  {journal} {\bibinfo  {journal} {Phys. Rev. B}\ }\textbf {\bibinfo {volume}
  {100}},\ \bibinfo {pages} {115132} (\bibinfo {year} {2019})}\BibitemShut
  {NoStop}%
\bibitem [{\citenamefont {Chowdhury}\ and\ \citenamefont
  {Berg}(2019)}]{chowdhury2019}%
  \BibitemOpen
  \bibfield  {author} {\bibinfo {author} {\bibfnamefont {D.}~\bibnamefont
  {Chowdhury}}\ and\ \bibinfo {author} {\bibfnamefont {E.}~\bibnamefont
  {Berg}},\ }\href {https://arxiv.org/abs/1908.02757} {\bibfield  {journal}
  {\bibinfo  {journal} {arXiv:1908.02757}\ } (\bibinfo {year}
  {2019})}\BibitemShut {NoStop}%
\bibitem [{\citenamefont {Hauck}\ \emph {et~al.}(2019)\citenamefont {Hauck},
  \citenamefont {Klug}, \citenamefont {Esterlis},\ and\ \citenamefont
  {Schmalian}}]{hauck2019}%
  \BibitemOpen
  \bibfield  {author} {\bibinfo {author} {\bibfnamefont {D.}~\bibnamefont
  {Hauck}}, \bibinfo {author} {\bibfnamefont {M.~J.}\ \bibnamefont {Klug}},
  \bibinfo {author} {\bibfnamefont {I.}~\bibnamefont {Esterlis}}, \ and\
  \bibinfo {author} {\bibfnamefont {J.}~\bibnamefont {Schmalian}},\ }\href
  {https://arxiv.org/abs/1911.04328} {\bibfield  {journal} {\bibinfo  {journal}
  {arXiv:1911.04328}\ } (\bibinfo {year} {2019})}\BibitemShut {NoStop}%
\bibitem [{\citenamefont {Wu}\ \emph {et~al.}(2019)\citenamefont {Wu},
  \citenamefont {Abanov}, \citenamefont {Wang},\ and\ \citenamefont
  {Chubukov}}]{wu2019}%
  \BibitemOpen
  \bibfield  {author} {\bibinfo {author} {\bibfnamefont {Y.-M.}\ \bibnamefont
  {Wu}}, \bibinfo {author} {\bibfnamefont {A.}~\bibnamefont {Abanov}}, \bibinfo
  {author} {\bibfnamefont {Y.}~\bibnamefont {Wang}}, \ and\ \bibinfo {author}
  {\bibfnamefont {A.~V.}\ \bibnamefont {Chubukov}},\ }\href {\doibase
  10.1103/PhysRevB.99.144512} {\bibfield  {journal} {\bibinfo  {journal} {Phys.
  Rev. B}\ }\textbf {\bibinfo {volume} {99}},\ \bibinfo {pages} {144512}
  (\bibinfo {year} {2019})}\BibitemShut {NoStop}%
\bibitem [{\citenamefont {Chubukov}\ \emph {et~al.}(2019)\citenamefont
  {Chubukov}, \citenamefont {Abanov}, \citenamefont {Wang},\ and\ \citenamefont
  {Wu}}]{chubukov2019}%
  \BibitemOpen
  \bibfield  {author} {\bibinfo {author} {\bibfnamefont {A.~V.}\ \bibnamefont
  {Chubukov}}, \bibinfo {author} {\bibfnamefont {A.}~\bibnamefont {Abanov}},
  \bibinfo {author} {\bibfnamefont {Y.}~\bibnamefont {Wang}}, \ and\ \bibinfo
  {author} {\bibfnamefont {Y.-M.}\ \bibnamefont {Wu}},\ }\href
  {https://arxiv.org/abs/1912.01797} {\bibfield  {journal} {\bibinfo  {journal}
  {arXiv:1912.01797}\ } (\bibinfo {year} {2019})}\BibitemShut {NoStop}%
\bibitem [{\citenamefont {Yang}\ \emph {et~al.}(2017)\citenamefont {Yang},
  \citenamefont {Sobota}, \citenamefont {He}, \citenamefont {Wang},
  \citenamefont {Leuenberger}, \citenamefont {Soifer}, \citenamefont
  {Hashimoto}, \citenamefont {Lu}, \citenamefont {Eisaki}, \citenamefont
  {Moritz}, \citenamefont {Devereaux}, \citenamefont {Kirchmann},\ and\
  \citenamefont {Shen}}]{yang2017}%
  \BibitemOpen
  \bibfield  {author} {\bibinfo {author} {\bibfnamefont {S.-L.}\ \bibnamefont
  {Yang}}, \bibinfo {author} {\bibfnamefont {J.~A.}\ \bibnamefont {Sobota}},
  \bibinfo {author} {\bibfnamefont {Y.}~\bibnamefont {He}}, \bibinfo {author}
  {\bibfnamefont {Y.}~\bibnamefont {Wang}}, \bibinfo {author} {\bibfnamefont
  {D.}~\bibnamefont {Leuenberger}}, \bibinfo {author} {\bibfnamefont
  {H.}~\bibnamefont {Soifer}}, \bibinfo {author} {\bibfnamefont
  {M.}~\bibnamefont {Hashimoto}}, \bibinfo {author} {\bibfnamefont {D.~H.}\
  \bibnamefont {Lu}}, \bibinfo {author} {\bibfnamefont {H.}~\bibnamefont
  {Eisaki}}, \bibinfo {author} {\bibfnamefont {B.}~\bibnamefont {Moritz}},
  \bibinfo {author} {\bibfnamefont {T.~P.}\ \bibnamefont {Devereaux}}, \bibinfo
  {author} {\bibfnamefont {P.~S.}\ \bibnamefont {Kirchmann}}, \ and\ \bibinfo
  {author} {\bibfnamefont {Z.-X.}\ \bibnamefont {Shen}},\ }\href {\doibase
  10.1103/PhysRevB.96.245112} {\bibfield  {journal} {\bibinfo  {journal} {Phys.
  Rev. B}\ }\textbf {\bibinfo {volume} {96}},\ \bibinfo {pages} {245112}
  (\bibinfo {year} {2017})}\BibitemShut {NoStop}%
\bibitem [{\citenamefont {Zaanen}(2019)}]{zaanen2019}%
  \BibitemOpen
  \bibfield  {author} {\bibinfo {author} {\bibfnamefont {J.}~\bibnamefont
  {Zaanen}},\ }\href {\doibase 10.21468/SciPostPhys.6.5.061} {\bibfield
  {journal} {\bibinfo  {journal} {SciPost Phys.}\ }\textbf {\bibinfo {volume}
  {6}},\ \bibinfo {pages} {61} (\bibinfo {year} {2019})}\BibitemShut {NoStop}%
\bibitem [{\citenamefont {J.~Zaanen}(2015)}]{Zaanen}%
  \BibitemOpen
  \bibfield  {author} {\bibinfo {author} {\bibfnamefont {Y.~W. S. K.~S.}\
  \bibnamefont {J.~Zaanen}, \bibfnamefont {UY.~Liu}},\ }\href@noop {} {\emph
  {\bibinfo {title} {Holographic duality in condensed matter physics}}}\
  (\bibinfo  {publisher} {Cambridge University Press},\ \bibinfo {year}
  {2015})\BibitemShut {NoStop}%
\bibitem [{\citenamefont {Paramekanti}\ \emph {et~al.}(1998)\citenamefont
  {Paramekanti}, \citenamefont {Trivedi},\ and\ \citenamefont
  {Randeria}}]{Paramekanti1998}%
  \BibitemOpen
  \bibfield  {author} {\bibinfo {author} {\bibfnamefont {A.}~\bibnamefont
  {Paramekanti}}, \bibinfo {author} {\bibfnamefont {N.}~\bibnamefont
  {Trivedi}}, \ and\ \bibinfo {author} {\bibfnamefont {M.}~\bibnamefont
  {Randeria}},\ }\href {\doibase 10.1103/PhysRevB.57.11639} {\bibfield
  {journal} {\bibinfo  {journal} {Phys. Rev. B}\ }\textbf {\bibinfo {volume}
  {57}},\ \bibinfo {pages} {11639} (\bibinfo {year} {1998})}\BibitemShut
  {NoStop}%
\end{thebibliography}
%

\acknowledgements
We thank Chandan Setty for extensive discussions on his work which ultimately motivated the search for an exactly solvable model with Mottness and superconductivity, G.\ La Nave for a critical analysis of the results, and Christian Boyd for many helpful discussions.
PWP thanks DMR19-19143 for partial funding of this project and EWH was supported by the Gordon and Betty Moore Foundation EPiQS Initiative through the grant GBMF 4305.

\clearpage

\section*{Supplementary Materials}

\subsection*{Pair binding instability with zero temperature Hatsugai-Kohmoto Gibbs state}

Given the Gibbs state $\rho = e^{-\beta H_\HK}/Z = \sum_n e^{-\beta E_n} \proj{n}$, we fix the purification on the doubled Hilbert space with $\cH_B \simeq \cH_A$
\begin{equation}
    \ket{\beta}
    = \sum_n e^{-\beta E_n/2} \ket{n} \ox \ket{n}
    \in \cH_A \ox \cH_B
\end{equation}
satisfying $\tr_B \proj{\beta} = \rho$, and restrict
\begin{equation}
    A\adj = \sum_{k \not\in \Omega_2} \alpha_k b_k\adj
\end{equation}
to the singly-occupied and unoccupied regions $\Omega_1$ and $\Omega_0$ of the Brillouin zone.
Now $\ket{\psi} = A\adj \ox \Id \ket{\oo}$ has overlap
\begin{align}
    \qave{\oo}{b_k \ox 1}{\psi}
    &= \tr_{A,B} \proj{\oo} (b_k A\adj \ox \Id) \\
    &= \tr_A \rho b_k A\adj = \ave{b_k A\adj} \\
    &=
    \begin{cases}
        0 & \text{if } k \in \Omega_2, \\
        \frac{1}{4} \alpha_k & \text{if } k \in \Omega_1, \\
        \alpha_k & \text{if } k \in \Omega_0.
    \end{cases}
\end{align}

With $[b_k, H]$ as before, for each $k \in \Omega_1, \Omega_0$ we have
\begin{align}
    &\quad C_k i \hbar \partial_t \alpha_k(t=0) \NN\\
    &= \qave{\oo}{e^{i(H\ox\Id+\Id\ox H')t} (b_k \ox \Id) e^{-i(H\ox\Id+\Id\ox H')t}}{\psi} \eval{t=0} \\
    &= \qave{\oo}{[b_k, H] \ox \Id}{\psi} \\
    &= \P{2 \xi_k + U \ave{n_{k\dn}+n_{-k\up}}} C_k \alpha_k \NN\\
    &\quad- \frac{g}{L^d} \ave{1-n_{k\up}-n_{-k\dn}} \B{ \sum_{q \in \Omega_1} \frac{1}{4} \alpha_q + \sum_{q \in \Omega_0} \alpha_q }
\end{align}
where $C_k=1/4$ if $k \in \Omega_1$ and $C_k=1$ if $k \in \Omega_0$.
Then taking $\alpha_k(t) = e^{-iEt/\hbar}\alpha_k(0)$ recovers the same consistency equation
\begin{align}
    1
    &= -\frac{g}{L^d} \sum_{k \in \Omega_1,\Omega_0} \frac{\ave{1-n_{k\up}-n_{-k\dn}}}{E - 2 \xi_k - U \ave{n_{k\dn}+n_{-k\up}}} \\
    &= -g\int_{\mu}^{W/2} d\epsilon \frac{\rho(\epsilon)}{E-2\epsilon+2\mu}
\end{align}
since the numerator vanishes in the singly-occupied region $\Omega_1$.

\subsection*{Dyson equation for pair susceptibility}

In order to relate the pair susceptibility
\begin{equation}
    \chi(i\nu_n) \equiv \frac{1}{L^d} \int_{0}^{\beta} d\tau\; e^{i\nu_n \tau} \ave{T \Delta(\tau) \Delta\adj}_{g}
\end{equation}
to the bare pair susceptibility, $\chi_0(i\nu_n)$ at $g=0$, we work in the interaction picture where
\begin{equation}
    \ave{T \Delta(\tau) \Delta\adj}_{g}
    = \frac{\ave{T S(\beta, 0) \Delta(\tau) \Delta\adj}_0}{\ave{T S(\beta,0)}_0}
\end{equation}
for $\Delta(\tau)$ evolved in the Heisenberg picture (under $H$) in $\ave{\cdots}_g$, and evolved in the interaction picture (under $H_\HK$) in $\ave{\cdots}_0$.
Here
\begin{equation}
    S(\beta,0)
    = e^{\beta H_\HK} e^{-\beta H}
    = T_{\tau_i} \exp g \int_{0}^{\beta} d\tau_1 H_p(\tau_1),
\end{equation}
for $H_p(\tau_1)$ evolved in the interaction picture.
Then the numerator takes the form of a power series in $g$,
\begin{align}
    &\quad \ave{T S(\beta, 0) \Delta(\tau) \Delta\adj}_0 \NN\\
    &= \sum_{m=0}^{\oo} \frac{g^m}{m!} \ave{T \P{\int_{0}^{\beta} d\tau_1 H_p(\tau_1)}^{m} \Delta(\tau) \Delta\adj}_0
\end{align}
where each term
\begin{widetext}
    \begin{equation}
        \frac{1}{L^d} \ave{T \P{\int_{0}^{\beta} d\tau_1 H_p(\tau_1)}^m \Delta(\tau) \Delta\adj}_0 \\
        = \frac{1}{(L^d)^{m+1}} \idotsint_{0}^{\beta} d\tau_1 \cdots d\tau_m \ave{T \Delta\adj(\tau_1) \Delta(\tau_1) \cdots \Delta\adj(\tau_m) \Delta(\tau_m) \Delta(\tau) \Delta\adj}_0
    \end{equation}
\end{widetext}
factorizes, as $\chi_0$ does in Eq.~(\ref{eq:bare-factorization}), because each pair annihilator $b_k(\tau)$ in the sum
\begin{equation}
    \Delta(\tau)
    = \sum_k b_k(\tau)
    = \sum_k e^{-\tau(2\xi_k + U (n_{k\dn} + n_{-k\up} - 1))} b_k
\end{equation}
evolves (in the interaction picture under $H_\HK$) as a multiple of an unevolved pair annihilator $b_k$.
The denominator then removes all disconnected factorizations with either $\ave{T \Delta\adj(\tau_i) \Delta(\tau_i)}_0$ for any time $\tau_i$ or $\ave{T \Delta(\tau) \Delta\adj}_0$.
This leaves only those factorizations with the form of an $m$-wise convolution, resulting in
\begin{equation}
    \chi(i\nu_n)
    = \sum_{m=1}^{\oo} g^{m-1} (\chi_0(i\nu_n))^m
    = \frac{\chi_0(i\nu_n)}{1 - g \chi_0(i\nu_n)}.
\end{equation}

\subsection*{Example of $T_c$ and $\Delta$ calculation}

To ease calculations, consider $\epsilon_k$ such that $\rho(\omega) = \frac{1}{L^d} \sum_k \delta(\omega - \epsilon_k) = \frac{1}{W}$ for $-\frac{W}{2} < \omega < \frac{W}{2}$. We will focus on the half-filled metal i.e. $U < W$ and $\mu = U/2$.

The single-particle density of states (DOS) can be broken into contributions from different regions of momentum space
\begin{align}
N(\omega) &= N_0(\omega) + N_2(\omega) + \frac{1}{2} N_1(\omega) \\
N_0(\omega) &= \theta(\omega) \rho(\omega + U/2) \\
N_2(\omega) &= \theta(-\omega) \rho(\omega - U/2) \\
N_1(\omega) &= \theta(-\omega)\theta(\omega+U) \rho(\omega + U/2) \\
&+ \theta(\omega)\theta(-\omega+U) \rho(\omega - U/2)
\end{align}
The effective DOS for calculating $T_c$ and $\Delta$ are
\begin{align}
N'(\omega) &= N_0(\omega) + N_2(\omega) + \frac{1}{4} N_1(\omega) \\
N''(\omega) &= N_0(\omega) + N_2(\omega) + N_1(\omega) \\
\end{align}

\paragraph{Superconducting temperature $T_c$.}

The susceptibility diverges when
\begin{equation}
\frac{1}{g} = \chi_0(0) = \int d\omega \; N'(\omega) \frac{\tanh{\frac{\beta\omega}{2}}}{2 \omega}.
\end{equation}
Set $x = \beta\omega/2$ and integrate by parts
\begin{align}
\frac{1}{g} = -\frac{1}{2} \int \dd{x} \ln x \bigg[ & N'(\frac{2 x}{\beta}) \sech^2 x \\
& + \qty(\dv{x} N'\qty(\frac{2 x}{\beta})) \tanh x \bigg].
\end{align}
For $T \ll U, W$ this becomes
\begin{align}
\frac{1}{g} &= \frac{1}{W} \ln{\frac{\beta(W-U)}{4}} + \frac{1/4}{W} \ln{\frac{\beta U}{4}} \\
&- N'(0)\qty(-\ln(\frac{4}{\pi}) - \gamma) \\
\frac{W}{g} &= \ln{\frac{\beta(W-U)}{4}} + \frac{1}{4} \ln{\frac{\beta U}{4}} - \frac{5}{4}\qty(-\ln(\frac{4}{\pi}) - \gamma)
\end{align}
where $\gamma \approx 0.577$ is Euler's constant. The solution gives the transition temperature
\begin{equation}
T_c = \qty(W-U)^{4/5} U^{1/5} \frac{e^\gamma}{\pi} e^{-\frac{4}{5} \frac{W}{g}}.
\end{equation}

\paragraph{Superconducting gap $\Delta$.}

The gap equation is given by
\begin{align}
1 &= \frac{g}{2} \int \dd{\omega} \frac{N''(\omega)}{\sqrt{\omega^2 + \abs{\Delta}^2}} \\
&= \frac{g}{W} \sinh[-1](\frac{W-U}{2\Delta}) + \frac{\alpha g}{W} \sinh[-1](\frac{U}{2\Delta})
\end{align}
For $\Delta \ll U, W$ this becomes
\begin{align}
\frac{1}{g} = \frac{1}{W} \ln(\frac{W-U}{\Delta}) + \frac{\alpha}{W} \ln(\frac{U}{\Delta})
\end{align}
which can be solved to find
\begin{equation}
\Delta = (W-U)^{1/2} U^{1/2} e^{-\frac{W}{2 g}}
\end{equation}

\subsection*{Variational ground state}

Consider the variational wave function
\begin{align}
\ket{\psi} = \prod_{k > 0} \qty(x_k + y_k b_k^\dagger b_{-k}^\dagger + \frac{z_k}{\sqrt{2}} \qty(b_k^\dagger + b_{-k}^\dagger)) \ket{0}.
\end{align}
$\braket{\psi}{\psi} = 1$ is satisfied if $\abs{x_k}^2 + \abs{y_k}^2 + \abs{z_k}^2 = 1$. This generalizes the BCS wavefunction, which corresponds to $x_k = u_k^2$, $y_k = v_k^2$, $z_k = \sqrt{2} u_k v_k$. Furthermore, the state defined by $x_k = 1$ for $k \in \Omega_0$, $z_k = 1$ for $k \in \Omega_1$, and $y_k = 1$ for $k \in \Omega_2$ is a ground state of the HK model. Note that although one signal of pair condensation in the BCS wavefunction is the presence of nonzero $u_k v_k \propto z_k$, this state is not a pair condensate.

In the free fermion case, the ground state in the absence of pairing is the filled Fermi sea, with $u_k = 1$ for $k \in \Omega_0$ and $v_k = 1$ for $k \in \Omega_2$. For a small pairing interaction $g$, the variational ground state with pairing is very similar but with both $u_k$ and $v_k$ non-zero near the boundary of $\Omega_0$ and $\Omega_2$, namely the Fermi surface. In the HK model with weak pairing ($g \ll U, W$), we similarly expect that both $x_k$ and $z_k$ become nonzero near the boundary of $\Omega_0$ and $\Omega_1$ and both $y_k$ and $z_k$ become nonzero near the boundary of $\Omega_1$ and $\Omega_2$.

Again we try to minimize $\expval{H}{\psi}$. For all $k > 0$, $p > 0$, and $k \neq p$,
\begin{align}
\expval{n_{k\sigma}}{\psi} &= \abs{y_k}^2 + \frac{\abs{z_k}^2}{2} \\
\expval{n_{k\uparrow}n_{k\downarrow}}{\psi} &= \abs{y_k}^2\\
\expval{b_k^\dagger b_{k}}{\psi} &= \abs{y_k}^2 + \frac{\abs{z_k}^2}{2} \\
\expval{b_k^\dagger b_{-k}}{\psi} &= \frac{\abs{z_k}^2}{2} \\
\expval{b_k}{\psi} &= \frac{1}{\sqrt{2}} \qty(x_k^*z_k + z_k^*y_k) \\
\expval{b_k^\dagger b_{p}}{\psi} &= \frac{1}{2}\qty(z_k^*x_k + y_k^*z_k)\qty(x_p^*z_p + z_p^*y_p).
\end{align}
The same equations apply if we take $k \to -k$, $p \to -p$ on the left hand sides. Combining everything, and ignoring terms like $g' \sum_k \dots$ that do not scale extensively in the thermodynamic limit,
\begin{align}
\expval{H}{\psi} &= \sum_{k > 0} \xi_k \qty(4 \abs{y_k}^2 + 2 \abs{z_k}^2) + U \qty(2 \abs{y_k}^2) \\
&\quad -g' \sum_{k, p > 0; k\neq p} 2 \qty(z_k^*x_k + y_k^*z_k)\qty(x_p^*z_p + z_p^*y_p) \\
&= \sum_{k > 0} \qty(4 \xi_k + 2 U) \abs{y_k}^2 + 2 \xi_k \abs{z_k}^2 \\
&\quad - 2 g' \sum_{k, p > 0; k\neq p} \qty(z_k^*x_k + y_k^*z_k)\qty(x_p^*z_p + z_p^*y_p)
\end{align}
For each $k$, introduce a lagrange multiplier $\lambda_k$ to enforce normalization.
\begin{align}
0 &= \pdv{x_k} \qty[\expval{H}{\psi} + \lambda_k \qty(\abs{x_k}^2 + \abs{y_k}^2 + \abs{z_k}^2 -1)] \\
&= \lambda_k x_k^* - 2 g' z_k^*\sum_{p>0, p\neq k} \qty(x_p^*z_p + z_p^*y_p) \\
\lambda_k &= 2 \frac{z_k^*}{x_k^*} O,
\end{align}
where $O = g' \sum_{p>0} \qty(x_p^*z_p + z_p^*y_p)$ (now including the contribution $p=k$, which is a $\order{1/L^d}$ difference).
\begin{align}
0 = \pdv{y_k^*} \qty[\dots] &= \qty(4 \xi_k + 2 U) y_k - 2 z_k O + \lambda_k y_k \\
&= \qty(4 \xi_k + 2 U) y_k - 2\qty(z_k - \frac{z_k^* y_k}{x_k^*}) O \\
2 \xi_k + U &= \qty(\frac{z_k}{y_k} - \frac{z_k^*}{x_k^*}) O \label{eqy}
\end{align}
\begin{align}
0 = \pdv{z_k^*} \qty[\dots] &= 2\xi_k z_k - 2\qty(x_k O + y_k O^*) + \lambda_k z_k \\
&= 2\xi_k z_k - 2\qty(x_k O + y_k O^* - \frac{\abs{z_k}^2}{x_k^*} O) \\
\xi_k x_k^* z_k &= \qty(\abs{x_k}^2 - \abs{z_k}^2) O + x_k^* y_k O^* \\
\xi_k &= \qty(\frac{x_k}{z_k} - \frac{z_k^*}{x_k^*}) O + \frac{y_k}{z_k} O^*. \label{eqlower}
\end{align}
In the last lines, we take the limit $L^d \to \infty$, so we ignore the $g'$ on the LHS of and also replace the sum in $O$ with a sum over all momentum. Subtracting (\ref{eqlower}) from (\ref{eqy}) gives
\begin{equation}
\xi_k + U = \qty(\frac{z_k}{y_k} - \frac{x_k}{z_k}) O - \frac{y_k}{z_k} O^*. \label{equpper}
\end{equation}
Using $\xi_k^l = \xi_k$ and $\xi_k^u = \xi_k + U$, and assuming everything is real,
\begin{align}
\xi_k^l &= \qty(\frac{x_k}{z_k} + \frac{y_k}{z_k} - \frac{z_k}{x_k}) O \\
\xi_k^u &= -\qty(\frac{x_k}{z_k} + \frac{y_k}{z_k} - \frac{z_k}{y_k}) O.
\end{align}
It is straightforward to check for $U = 0$, combining these equations produces exactly the BCS result, even though we started with a more general wavefunction. This system of two equations is possible to solve analytically, but requires finding the roots of a quartic equation.

\paragraph{Weak coupling $g \ll U$ and $g \ll W$.}

First, rewrite
\begin{align}
\xi_k^l x_k z_k &= \qty(x_k^2 - z_k^2 + x_k y_k) O \label{gap_eq_xz} \\
\xi_k^u y_k z_k &= \qty(z_k^2 - y_k^2 - x_k y_k) O \label{gap_eq_yz},
\end{align}
which now looks very similar to the BCS case, apart from the $x_k y_k$ terms.

If $g \ll U$, we expect that there are still well defined regions $\Omega_0$, $\Omega_1$, $\Omega_2$, such that mixing occurs only between $x_k$ and $z_k$ or between $y_k$ and $z_k$ and never $x_k$ and $y_k$. Is it safe to drop the $x_k y_k$ terms from (\ref{gap_eq_xz}) and (\ref{gap_eq_yz})? Consider a $k$ point where $\xi_k^l < \xi_k^u < 0$. Here we expect $1 \approx y_k \gg z_k \gg x_k$. If $x_k$, $y_k$, $z_k$ are all positive, (\ref{gap_eq_xz}) can only be satisfied if $z_k^2 \gg x_k y_k$. A similar argument can be made for (\ref{gap_eq_yz}). Therefore we drop $x_k y_k$ from both equations and work with
\begin{align}
\xi_k^l x_k z_k &= \qty(x_k^2 - z_k^2) O \\
\xi_k^u y_k z_k &= \qty(z_k^2 - y_k^2) O.
\end{align}
Change variables to
\begin{align}
x_k^2 - z_k^2 &= \frac{\xi_k^l}{E_k^l}\qty(1-y_k^2), & 2 x_k z_k &= \frac{\Delta_k^l}{E_k^l}\qty(1-y_k^2) \\
E_k^l &= \sqrt{{\xi_k^l}^2 + {\Delta_k^l}^2} \\
z_k^2 - y_k^2 &= \frac{\xi_k^u}{E_k^u}\qty(1-x_k^2), & 2 y_k z_k &= \frac{\Delta_k^u}{F_k^u}\qty(1-x_k^2) \\
E_k^u &= \sqrt{{\xi_k^u}^2 + {\Delta_k^u}^2}
\end{align}
to get
\begin{align}
\Delta_k^l &= g' \sum_{p>0} \frac{\Delta_p^l}{E_p^l} \qty(1-y_k^2) + \frac{\Delta_p^u}{E_p^u} \qty(1-x_k^2) \\
\Delta_k^u &= g' \sum_{p>0} \frac{\Delta_p^l}{E_p^l} \qty(1-y_k^2) + \frac{\Delta_p^u}{E_p^u} \qty(1-x_k^2)
\end{align}
from which we see that there is only a single momentum-independent parameter $\Delta$ defined by
\begin{align}
1 &= g' \sum_{k>0} \frac{1-y_k^2}{\sqrt{{\xi_k^l}^2 + \Delta^2}}  + \frac{1-x_k^2}{\sqrt{{\xi_k^u}^2 + \Delta^2}} \\
1 &= \frac{g}{2} \int \dd{\omega} \frac{N''(\omega)}{\sqrt{\omega^2 + \Delta^2}}.
\end{align}
This is the same as the BCS gap equation, but with an effective density of states
\begin{equation}
N''(\omega) = \frac{1}{L^d} \sum_k \delta(\omega - \xi_k^l) (1-y_k^2) + \delta(\omega - \xi_k^u) (1-x_k^2)
\end{equation}
Because we are considering $g \ll U$, to a very good approximation $1-y_k^2 = \theta(\xi_k^u)$ and $1-x_k^2 = \theta(-\xi_k^l)$.
\begin{equation}
N''(\omega) = \frac{1}{L^d} \sum_k \delta(\omega - \xi_k^l) \theta(\xi_k^u) + \delta(\omega - \xi_k^u) \theta(-\xi_k^l) \label{effDOS}
\end{equation}
Note that $N''(\omega)$ is \emph{not} the single-particle density of states of the HK model. In fact it is larger than or equal to it for all $\omega$, and $\int \dd{\omega} N''(\omega) \geq 1$. (\ref{effDOS}) may be rewritten as
\begin{align}
N''(\omega) &= \sum_{k \in \Omega_0} \delta(\omega-\xi_k^l) + \sum_{k \in \Omega_2} \delta(\omega-\xi_k^u) \\
&+ \sum_{k \in \Omega_1} \delta(\omega-\xi_k^l) + \delta(\omega-\xi_k^u).
\end{align}

\subsection*{Mean-field calculation of spectral function}
To calculate the spectral function in Fig.~\ref{fig:phase-diagram}, we treat the pairing interaction at the mean-field level and the Mott ($U$) interaction exactly.
\begin{align}
H_{pair} &= - g' \sum_{k k'} b_k^\dagger b_{k'} \\
&= - g' \sum_{k k'} \expval{b_k^\dagger} b_{k'} + b_k^\dagger \expval{b_{k'}} \\
&\quad+ \qty(b_k^\dagger - \expval{b_k^\dagger})\qty(b_{k'} - \expval{b_{k'}}) - \expval{b_k^\dagger}\expval{b_{k'}}.
\end{align}
In the mean-field approximation, the third term (quadratic in fluctuations) is dropped. The last term is an inconsequential constant. Define
\begin{align}
\Delta &= -g' \sum_k \expval{b_k} \label{eq:MFTDelta} \\
H_{pair}^{MF} &= \sum_k \Delta^* b_k + \Delta b_k^\dagger
\end{align}
The full Hamiltonian becomes
\begin{equation}
H^{MF} = \sum_k \xi_k \qty(n_{k\uparrow} + n_{k\downarrow}) + U  n_{k\uparrow} n_{k\downarrow} + \Delta^* b_k + \Delta b_k^\dagger.
\end{equation}
Here, the pairing is treated at the mean-field level but the Mott interaction is treated exactly. While the Hamiltonian no longer separates completely in $k$-space, each $k$ couples only to $-k$.
\begin{align}
H^{MF} &= \sum_{k>0} H_k^{MF} \\
H_k^{MF} &= \xi_k \qty(n_{k\uparrow} + n_{k\downarrow} + n_{-k\uparrow} + n_{-k\downarrow}) \\
&\quad+ U \qty(n_{k\uparrow} n_{k\downarrow} + n_{-k\uparrow} n_{-k\downarrow}) \\
&\quad + \Delta^* \qty(b_k + b_{-k}) + \Delta \qty(b_k^\dagger + b_{-k}^\dagger).
\end{align}
$H^{MF}$ may be solved by exact diagonalization of each $H_k^{MF}$, yielding 16 eigenstates and energies for each $k$. $\Delta$ is adjusted for self-consistency such that Eq.~\ref{eq:MFTDelta} is satisfied. The single-particle spectral function is calculated directly from its spectral representation
\begin{equation}
A(k,\omega) = \sum_{n m} \abs{\mel**{n}{c_k}{m}}^2 \qty(\rho_n + \rho_m) \delta(\omega + E_n - E_m),
\end{equation}
where $\ket{n}$ and $\ket{m}$ are eigenstates of $H_k^{MF}$, $\rho_n = e^{-\beta E_n}/Z$, and the partition function $Z = \sum_n e^{-\beta E_n}$.

\subsection*{Calculation of superfluid stiffness}
The superfluid stiffness can be calculated as
\begin{equation}
\frac{D_s}{\pi} = \frac{1}{L^d} \qty(\langle K_{x x} \rangle - \int_0^\beta d\tau \,\langle J_x(\tau) J_x \rangle),
\end{equation} \label{eq:stiff}
where
\begin{align}
K_{x x} &=  \sum_{k \sigma} \frac{\partial^2 \epsilon_k}{\partial k_x^2} c_{k\sigma}^\dagger c_{k\sigma} \\
J_x &= \sum_{k \sigma} \frac{\partial \epsilon_k}{\partial k_x} c_{k\sigma}^\dagger c_{k\sigma}.
\end{align}
As for the spectral function, all expectation values are calculated by exact diagonalization of $H_k^{MF}$. In Fig.~\ref{fig:stiff}, a $64\times64$ grid of $k$-points is used.

For a nearest neighbor tight-binding band structure as considered throughout this work, $\frac{\partial^2 \epsilon_k}{\partial k_x^2} = 2t \cos k_x$, in units where the lattice constant $a=1$. Therefore $\ev{K_{x x}}$ is simply the (negative) kinetic energy along the $x$-direction bonds. In the spectral representation, we see that
\begin{equation}
\int_0^\beta d\tau \,\langle J_x(\tau) J_x \rangle = \sum_{n m} \abs{\mel**{n}{J_x}{m}}^2 \frac{\rho_m - \rho_n}{E_n - E_m}
\end{equation}
is nonnegative, so $\frac{\pi}{L^d} \ev{K_{xx}}$ is an upper bound to the superfluid stiffness\cite{Paramekanti1998}.

\subsection*{Superconducting energy scales in $d=1,2,3$}

\begin{figure*}[h]
    \centering
    \includegraphics[width=\textwidth]{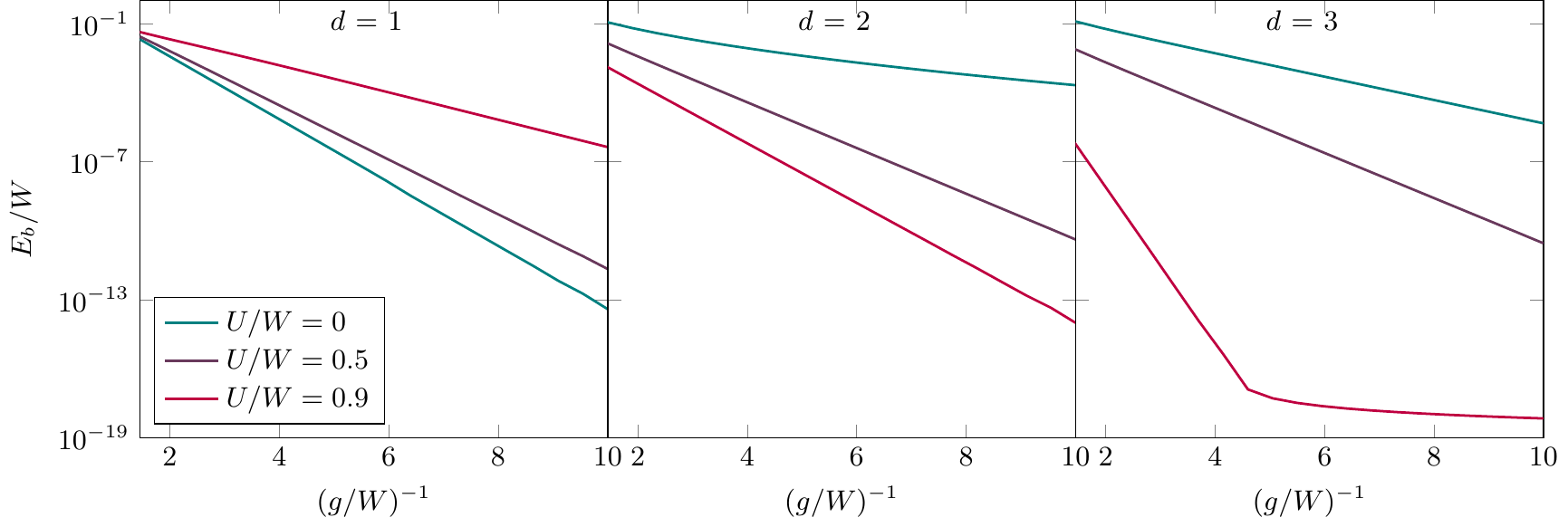}
    \caption{ Cooper pair binding energy $E_b$ at half-filling.}
\end{figure*}

\begin{figure*}[h]
    \centering
    \includegraphics[width=\textwidth]{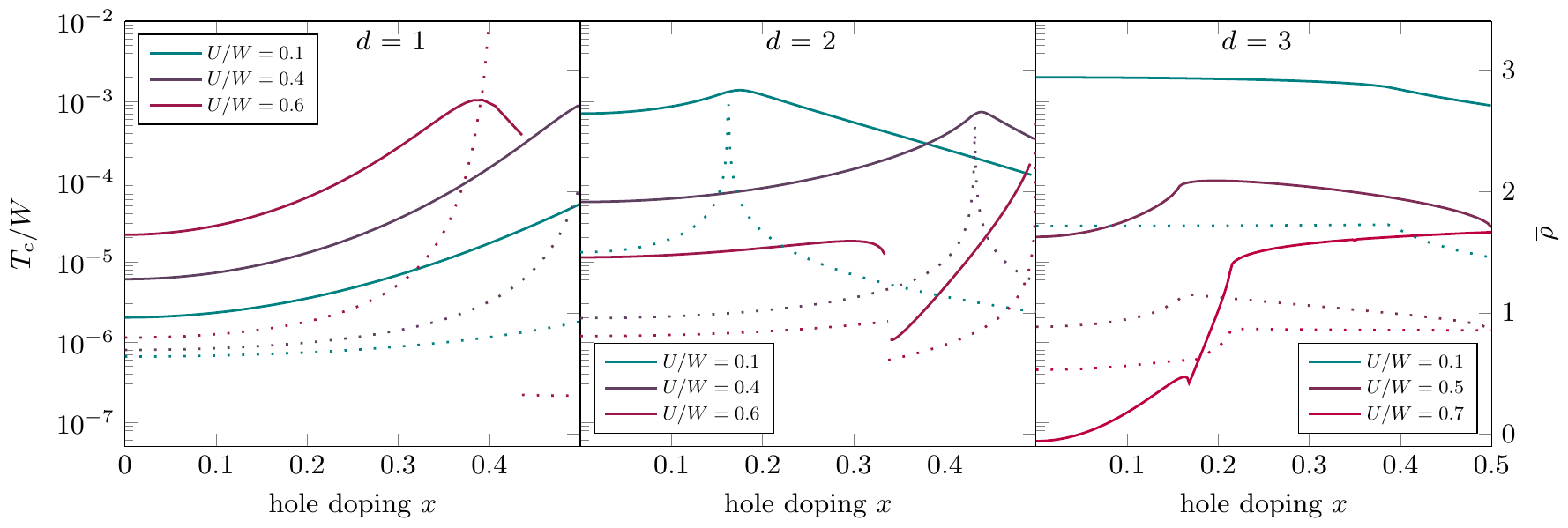}
    \caption{Superconducting temperature $T_c$ (solid) and mean density of states $\bar{\rho} = \half(\rho(\mu) + \rho(\mu-U))$ (dotted) at pair coupling $g/W=0.1$.}
\end{figure*}

\end{document}